# Pulse-Shape discrimination with the Counting Test Facility

H.O. Back[13a], M. Balata[1], G. Bellini[2], J. Benziger[3], S. Bonetti[2], B. Caccianiga[2], F. Calaprice[4], F. Dalnoki-Veress[4], D. D'Angelo[2], A de Bellefon[6], H. de Kerret[6], A. Derbin[7], A. Etenko[8], K. Fomenko[7], R. Ford[4], D. Franco[2], C. Galbiati[4], S. Gazzana[1], M.G. Giammarchi[2], M. Goeger[5], A. Goretti[4], C. Grieb[13], S. Hardy[13], A. Ianni[1], A. M. Ianni[4], G. Korga[1], Y. Kozlov[8], D. Kryn[6], M. Laubenstein[1], M. Leung[4], E. Litvinovich[8], P. Lombardi[2], L. Ludhova[2], I. Machulin[8], I. Manno[10], D. Manuzio[11], G. Manuzio[11], F. Masetti[12], K. McCarty[4], E. Meroni[2], L. Miramonti[2], M. Misiaszek[14], D. Montanari[1], M. E. Monzani[1b], V. Muratova[7], L. Niedermeier[5], L. Oberauer[5], M. Obolensky[6], F. Ortica[12], M. Pallavicini[11], L. Papp[1], L. Perasso[2], A. Pocar[4], R. S. Raghavan[13], G. Ranucci[2*], A. Razeto[1], A. Sabelnikov[1d], C. Salvo[11], S. Schoenert[9], T. Shutt[4c], H. Simgen[9], M. Skorokhvatov[8], O. Smirnov[7], A. Sonnenschein[4], A. Sotnikov[7], S. Sukhotin[8], Y. Suvorov[2], V. Tarasenkov[8], R. Tartaglia[1], G. Testera[11], D. Vignaud[6], R. B. Vogelaar[13], F. Von Feilitzsch[5], B. Williams[13], M. W´ojcik[14], O. Zaimidoroga[7], S. Zavatarelli[11] and G. Zuzel[9]

[1] INFN Laboratori Nazionali del Gran Sasso, SS 17 bis Km 18+910, I-67010 Assergi(AQ), Italy

[2] Dipartimento di Fisica Università and INFN, Milano, Via Celoria, 16 I-20133 Milano, Italy

[3] Department of Chemical Engineering, A-217 Engineering Quadrangle, Princeton NJ 08544-5263, USA

[4] Department of Physics, Princeton University, Jadwin Hall, Washington Rd, Princeton NJ 08544-0708, USA

[5] Technische Universität München, James Franck Strasse, E15 D-85747, Garching, Germany

[6] Astroparticule et Cosmologie APC, Collège de France, 11 place Marcelin Berthelot, 75231 Paris Cedex 05, France

[*] Corresponding author. Tel: +39-02-50317362; e-mail: gioacchino.ranucci@mi.infn.it.

[7] Joint Institute for Nuclear Research, 141980 Dubna, Russia

[8] RRC Kurchatov Institute, Kurchatov Sq.1, 123182 Moscow, Russia

[9] Max-Planck-Institut fuer Kernphysik,Postfach 103 980 D-69029, Heidelberg, Germany

[10] KFKI-RMKI, Konkoly Thege ut 29-33 H-1121 Budapest, Hungary

[11] Dipartimento di Fisica Università and INFN, Genova, Via Dodecaneso,33 I-16146 Genova, Italy

[12] Dipartimento di Chimica Università and INFN, Perugia, Via Elce di Sotto, 8 I-06123, Perugia, Italy

[13] Physics Department, Virginia Polytechnic Institute and State University, Robeson Hall, Blacksburg, VA 24061-0435, USA

[14] M.Smoluchowski Institute of Physics, Jagellonian University, PL-30059 Krakow, Poland

[a]Present address: Department of Physics North Carolina State University 2700 Stinson Dr, Box 8202, Raleigh, NC 27695

[b]Present address: Columbia University, Astrophysics Lab, 550 West 120th Street, New York NY 10027, USA

[c]Present address: Case Western Reserve University, Cleveland, Ohio 44118, USA

[c]On leave of absence from (8)

**Abstract**

Pulse shape discrimination (PSD) is one of the most distinctive features of liquid scintillators. Since the introduction of the scintillation technique in the field of particle detection, many studies have been carried out to characterize intrinsic properties of the most common liquid scintillator mixtures in this respect. Several application methods and algorithms able to achieve optimum discrimination performances have been developed. However, the vast majority of these studies have been performed on samples of small dimensions. The Counting Test Facility, prototype of the solar neutrino experiment Borexino, as a 4 ton spherical scintillation detector immersed in 1000 tons of shielding water, represents a unique opportunity to extend the small-sample PSD studies to a large-volume setup. Specifically, in this work we consider two different liquid scintillation mixtures employed in CTF, illustrating for both the PSD characterization results obtained either with the processing of the scintillation waveform through the optimum Gatti's method, or via a more conventional discrimination approach based on the charge content of the scintillation tail. The outcomes of this study, while interesting per se, are also of paramount importance in view of the expected Borexino detector performances, where PSD will be an essential tool in the framework of the background rejection strategy needed to achieve the required sensitivity to the solar neutrino signals.

## 1. Introduction

The achievement and the measurement of very low background levels in a large volume of liquid scintillator is the main goal of the Counting Test Facility (CTF), currently running underground at the Laboratory of Gran Sasso (LNGS) [1]. The CTF was constructed as a prototype of the Borexino detector, a real-time liquid scintillator detector mainly devoted to low energy solar neutrino physics [2]. The application of the pulse shape discrimination (PSD) to $\alpha$ and $\beta$ induced events in the scintillation medium is part of the program aimed at disentangling the decay events produced by the natural radioactive chains (i.e. $^{238}$U and $^{232}$Th) from the data set. In this work, the PSD separation capabilities measured in CTF are presented, as obtained in different data taking periods and conditions of the detector.

Two scintillation solutions have been tested: the first is the scintillator used in the low-energy solar neutrino detector Borexino and consists of a binary solution of 1,2,4-trimethylbenzene (pseudocumene, PC) as solvent and of 2,5-diphenyloxazole (PPO) as scintillating solute at a concentration of 1.5 g/l; the second is a “back-up solution” consisting of 1,2-dimethyl-4-(1-phenylethyl)-benzene (phenyl-o-xylylethane, PXE) as solvent, 1,4-diphenylbenzene (para-Terphenyl, p-Tp) as a primary solute (2.0 g/l), and 1,4-bis(2-methylstyryl)benzene (bis-MSB) (20 mg/l) as a secondary wavelength-shifter solute.

The PC sum formula is $C_9H_{12}$ and the molecular weight is 120.2 g/mol. It has a density of 0.876 $g/cm^3$ and a flash point of 48 $^oC$. The PC solvent was provided by the Italian company Enichem (now Polimeri Europa) located in Sardinia.

The PXE has a sum formula $C_{16}H_{18}$ with a molecular weight of 210.3 g/mol. It has a high density (0.988 $g/cm^3$), a low vapor pressure, and a high flash point of $145^oC$. The PXE used in this work has been produced by the American Koch Special Chemical Company in Texas.

CTF is an un-segmented transparent nylon sphere filled with 4.8 $m^3$ of liquid scintillator, with a graded shielding against external background (see Fig. 1). A complete description can be found in [1] [3].

The active volume is immersed in high-purity water that works as shielding against the ambient gamma rays and neutrons at the underground site. The three-dimensional light collection system is composed of one hundred photomultipliers equipped with light concentrators, mounted on an open structure around the active volume, and immersed in the water buffer. A light yield of about 350 photoelectrons per MeV of deposited energy in the scintillator is obtained. For each event, the collected charges and photon arrival times are recorded, allowing determination of the total deposited energy and the spatial position of each event in an off-line analysis. To record fast time-correlated events, each electronic processing channel has a duplicate auxiliary channel to process any following event that may occur within a time window of 8 ms. For the specific purpose of PSD, the PMT output signals are also summed and time integrated over the total pulse and the pulse tail, or digitized by the special Digital Processing Sampling Board (DPSA) described in **§4.2**.

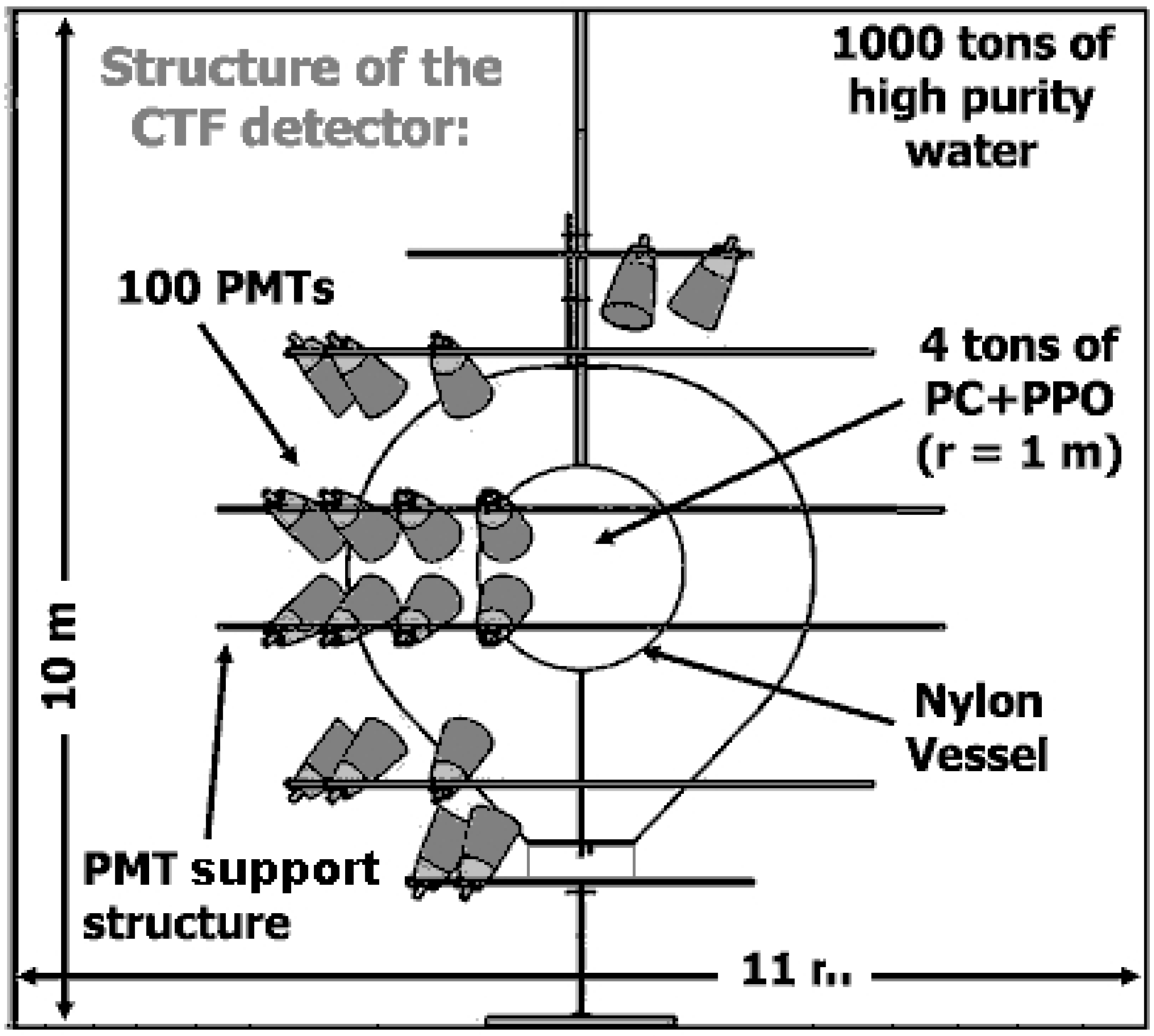

*Fig. 1 - Structure of the CTF detector.*

It is well known that the scintillator particle discrimination properties discussed in this paper are a peculiar feature of the scintillation pulse wave shapes induced by different interacting particles. The de-excitation of the scintillator which follows an interaction depends on many factors, with a special role played by the energy loss density; in particular, the larger the dE/dx, the greater is the slow component (i.e., the component that decays over long times) of the scintillation decay curve with respect to the fast component (that is, the component that decays more rapidly). This dependence upon dE/dx automatically provides a tool to distinguish particles of different type [4], and in particular in the case of our application, $\alpha$ events from $\beta$ signals.

## 2. Description of the discrimination methods

Given the intrinsic distinction between alpha and beta pulses due to the processes associated with the scintillation mechanism, several techniques and implementations have become of common use in the area of the liquid scintillation applications to exploit such difference in practical experimental situations. In this work we have focused our efforts towards two such approaches: the “tail to total” method (also known as charge integration method) and the so called “optimum filter”. In this section we briefly describe the main features of both implementations.

### *2.1. Tail to total (or charge) method*

The tail to total (or charge) method to accomplish PSD exploits, in a straightforward way, the above-mentioned dependence of the fast and slow portions of the scintillation pulse upon the dE/dx of the interacting particle. Its implementation, indeed, relies on the integration of the signal over two different intervals, one encompassing the entire pulse duration and the other only the tail: the ratio of the charge in the tail to that comprised in the whole signal is then adopted as the separation parameter. Recall that for heavily ionizing particles (like $\alpha$’s) the relative amount of light in the tail is larger than for relativistic electrons; therefore, the tail-to-total ratio is naturally an effective distinguishing tool, assuming values that are distinctively higher for alpha particles with respect to the electron case.

The statistics governing this method are quite simple, being essentially driven by the Poisson fluctuation of the number of photoelectrons in the tail (practically reduced to a Gaussian distribution); furthermore, specific circuitry for its implementation is not required, it being sufficient to properly combine standard charge ADC's with timing and trigger units.

Its conceptual simplicity and its straightforward hardware implementation make this approach particularly attractive, thus explaining the popularity it has gained in many applications, despite the fact that in term of the degree of discrimination achievable, it is sub-optimal with respect to the optimum linear filter method (Gatti’s method) described in the following paragraphs.

*2.2. Optimum Gatti's method*

The optimum Gatti's method [5] requires the knowledge both of the average pulse shape of the signals produced by the particles of the two species to be identified and of the individual random waveforms generated in the single excitations.

Specifically, denoting with $\overline{\alpha}(t)$ and $\overline{\beta}(t)$ the average time function of the α and β current pulses at the output of the photomultiplier, and with $\alpha_i$ and $\beta_i$ the number of photoelectrons emitted for individual random waveforms in the arbitrary short time intervals $\Delta t_i$ into which the duration of the signals can be divided, in Gatti's formulation the particle identification parameter G is obtained through the following weighted sums, for α and β respectively:

$$G_\alpha = \sum_i P_i \beta_i \quad (1)$$

$$G_\beta = \sum_i P_i \alpha_i \quad (2)$$

where the weights $P_i$ are given by:

$$P_i = \frac{(\overline{\alpha_i} - \overline{\beta_i})}{(\overline{\alpha_i} + \overline{\beta_i})} \quad (3)$$

Following (1), the weights are positive for the time interval in which the normalized curve $\overline{\alpha}(t)$ has values higher than that of $\overline{\beta}(t)$, and negative elsewhere.

After computation of the $P_i$ factors, given an unknown signal $S(t)$ (corresponding to a number of photoelectrons $S_i$ emitted in the time intervals $\Delta t_i$), its likelihood to have been produced by an α or β excitation can be described through a discrimination parameter, computed as follows:

$$G = \sum_i P_i S_i \quad (4)$$

As demonstrated in the original Gatti paper, the G parameters are distributed around mean values which are positive for the alpha particles and negative for the beta signals, provided that the signals are normalized to the same area and that processing time is extended up to when the tail of the scintillation light has totally vanished to zero. If the processing time is kept shorter, which may happen for practical considerations, then the mean values $\overline{G_\alpha}$ and $\overline{G_\beta}$ , though still of opposite sign, will be not perfectly balanced.

At the time when the method was first proposed, it was very difficult to sample the signal S(t) with sufficient accuracy to compute the G parameter. This restriction, which for many years prevented the practical adoption of Gatti's methodology, has finally been overcome with the introduction of modern flash ADC technology. This technique is, indeed, quite suitable for providing the acquisition, via digital sampling, of all the waveforms of interest.

Prior to proceeding with the hardware development, we evaluated the improvement in the expected PSD performance attainable with the optimum method compared to a classic implementation of the tail-to-total method. For this purpose we performed a Monte Carlo evaluation based on the features of the PMT's and the PC+PPO Borexino scintillator. The results are shown in Fig. 2 in terms of the factor of merit D, defined as $D = \Delta S / \sqrt{\sigma_\beta^2 + \sigma_\alpha^2}$ , where $\Delta S$ is the distance between the peaks of the distributions of the two separation parameters, and $\sigma_\alpha$ and $\sigma_\beta$ are the respective variances. Since D is a measure of the distance of the separation spectra, better discrimination is represented by higher values of D. Thus, from Fig. 2 we infer the global superior performances enabled by the Gatti method over the tail-to-total ratio. The plot shows, in particular, the significant improvement predicted in the range of 150-250 photoelectrons, of paramount importance for Borexino as well as for the CTF, since it is in this range that most of the $\alpha$ background to be rejected is located.

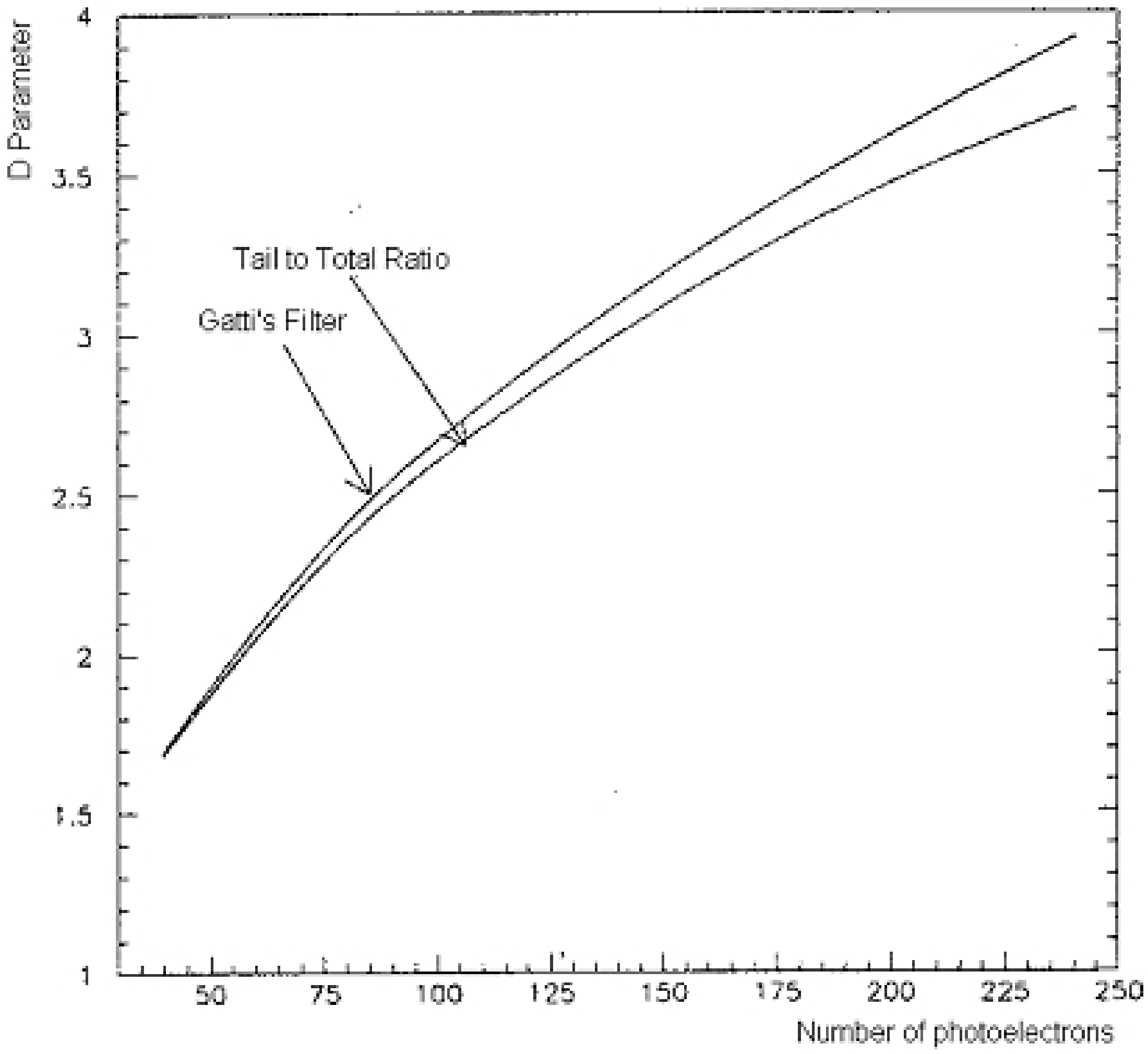

*Fig. 2 - This plot shows the superior performance, predicted via Monte Carlo, of Gatti's method compared to a conventional implementation of the charge (tail-to-total) method for pulse shape discrimination.*

### 3. Preliminary laboratory measurements

It is obvious that, irrespective to the adopted methodology, the ultimate discrimination achievable in a liquid scintillation detector depends upon the intrinsic light response under irradiation of the two species to be discriminated. For this reason, an important part of the R&D efforts connected with the Borexino project was devoted to the assessment of the intrinsic PSD discrimination power of the candidate scintillation cocktails through a set of small-scale laboratory experiments.

These laboratory investigations progressed along two different but complementary paths, one contemplating the quantitative determination of the average time profiles of the scintillation light for alpha and beta excitations [6], and the other focused on the direct application of the tail-to-total discrimination method to the samples under study [7]. Clearly the output of the former study is more complete, since the knowledge of the entire α and β light profiles allows one to evaluate (either numerically or by Monte Carlo) the discrimination features stemming from the application of various PSD algorithms to the α and β curves; nevertheless the latter investigation, providing a direct measure of the discrimination capability under the specific tail-to-total PSD method, not only represents a further confirmation of the results obtained in the previous measurements, but helps to shed light on the practical aspects of its actual implementation, especially in view of the successive application in the CTF detector. In the following we summarize briefly the results obtained from these studies.

### *3.1. Determination of the average time profile*

The determination of the time profile has been carried out exploiting the so-called "single-photon sampling technique" [8], which allows one to derive the scintillation decay curve statistically. The experimental set-up is shown in Fig. 3; it consists of two photomultipliers observing the scintillator specimen under test. One tube, closely coupled to the scintillator, detects all the occurring events, thus providing the zero-time reference. Instead, the other phototube is weakly coupled to the sample, via a proper neutral filter; in this way when it detects a signal, this is almost always due to a single photoelectron, and only very rarely to two or more photoelectrons. When this condition is fulfilled, the experimental "single-photon sampling technique" histogram faithfully reproduces, after the tube response is deconvolved, the original time profile of the light emission in the scintillator. This histogram is obtained by plotting the time difference between the zero reference signal and the time of emission of the anode signal of the low level tube, measured by the TDC in each event for which the detection of at least one photoelectron occurs.

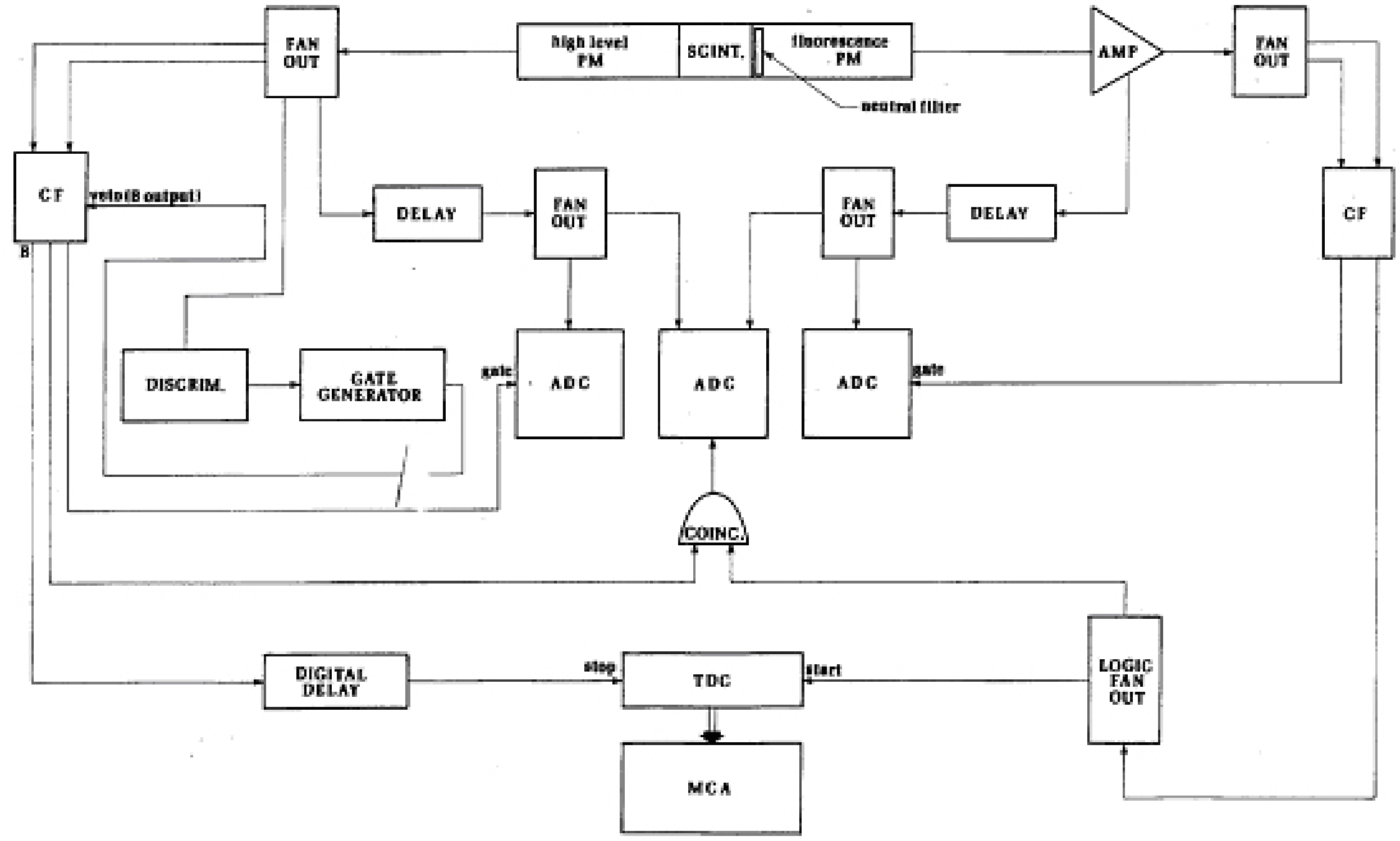

*Fig. 3 - Schematic block diagram of the experimental setup for the application of the single photon sampling technique.*

As shown in Fig. 3, the reference signal coming from the high level photomultiplier is routed to a CAMAC constant fraction discriminator whose output, delayed by means of a digital delay, is used to feed the stop input of a time-to-digital converter. The output of the low level phototube, also often referred to as the fluorescence photomultiplier, is amplified, discriminated by a constant fraction discriminator, and then used as input for the start of the TDC. This reverse configuration has been adopted to avoid the high rate of start signals not followed by any stop that would occur in the direct configuration, in which the start to the TDC is derived from the anode signal of the high level tube. In order to reduce the rate at the stop input of the TDC, a gate generator has been used to veto the constant fraction discriminator for a period of 8 μs after the occurrence of a pulse. This rate reduction was needed to avoid distortions of the measured curve arising from pile-up.

It is important to underline that, since the scintillation time decay is strongly affected by the presence of oxygen dissolved in the liquid, all the measured samples have been preliminarily subjected to a thorough purging with nitrogen.

In Fig. 4, as an example, we report the results of the measurement of the PC+PPO scintillator; in this figure, the responses of the scintillator to $\alpha$ and $\beta$ excitations (obtained with $^{210}$Po and $^{60}$Co sources) are displayed together, showing clearly, at least qualitatively, that such a scintillator features good discrimination capabilities.

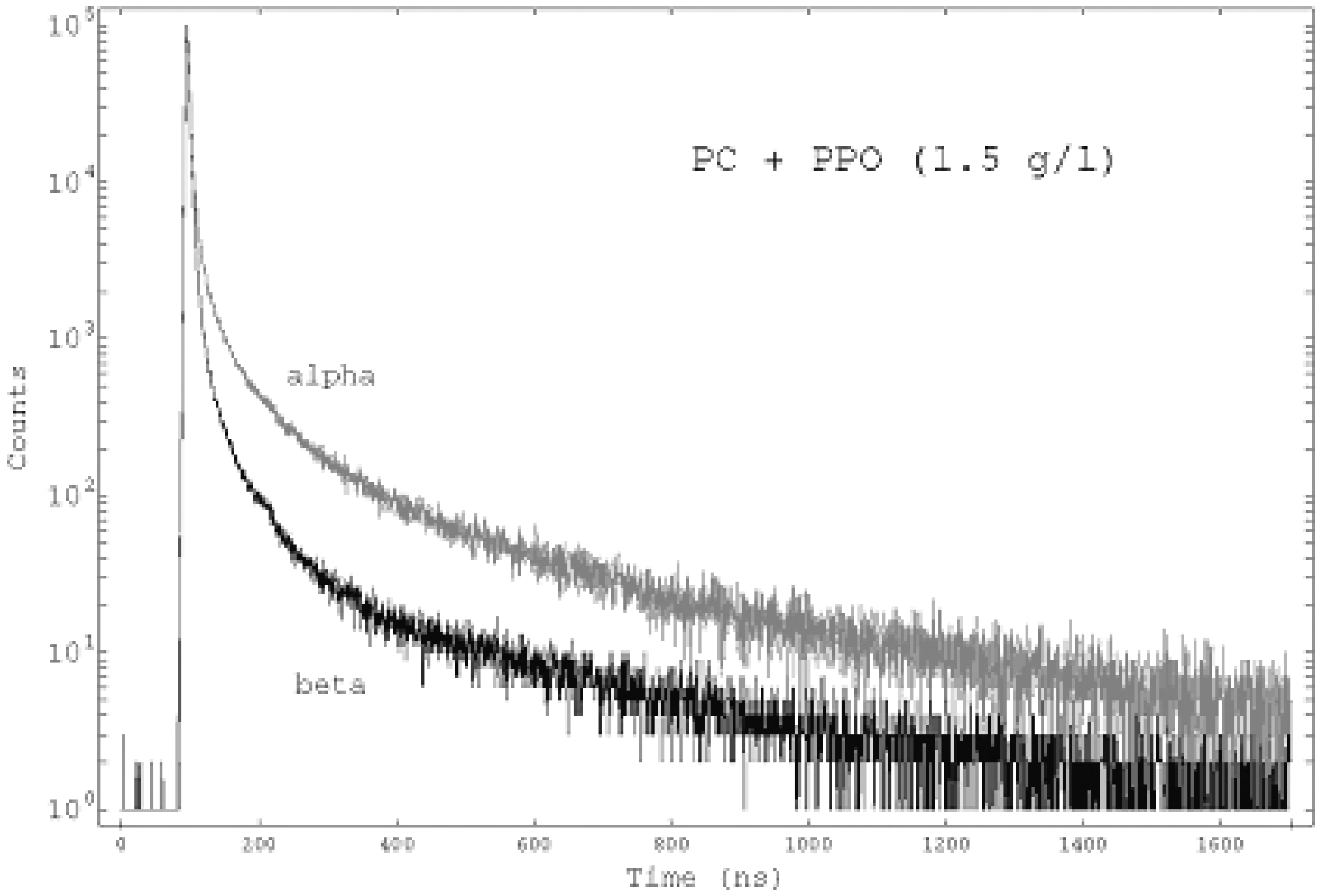

*Fig. 4 - Average time profiles of the scintillation pulses emitted by a PC+PPO (1.5 g/l) mixture under alpha and beta irradiation.*

Quantitatively, the light curves can be described, after deconvolving the tube response, as a weighted sum of exponential terms, i.e. $\sum_{i}^{n} \frac{q_i}{\tau_i} e^{-\frac{t}{\tau_i}}$ ; the values of the $q$ and $\tau$ parameters obtained in the measurements carried out for both the PC and PXE based scintillators are reported in Table 1. Since $q_1$ describes the amount of light in the fast component of the scintillation pulse, while $q_2$, $q_3$ and $q_4$ cumulatively represent the light in the slow component, from the table we infer that for both scintillators there is a distinctive, sizeable, decrease of light in the former component, and a corresponding increase in the latter, while passing from the beta to the alpha excitation. Thus, both mixtures are in principle suited to be used for α/β discrimination applications. The discussion of the comparison of the respective discrimination power is deferred to the illustration of the results obtained in the CTF detector in **§6.3**; it can, however, be noted from these measurements that the PXE scintillator features more light, under β irradiation, in the tail with respect to the PC scintillator, from which we can already infer that likely the PXE based mixture should exhibit a lower intrinsic discrimination capability.

| | Excit. | $\tau_1$ (ns) | $\tau_2$(ns) | $\tau_3$ (ns) | $\tau_4$ (ns) | $q_1$ | $q_2$ | $q_3$ | $q_4$ |
|---|---|---|---|---|---|---|---|---|---|
| PC/ PPO(1.5 g/l) | β | 3.57 | 17.61 | 59.50 | | 0.895 | 0.063 | 0.042 | |
| PC/ PPO(1.5 g/l) | α | 2.19 | 12.02 | 56.13 | 433.6 | 0.636 | 0.153 | 0.104 | 0.107 |
| PXE/p-TP(3.0 g/l)/ bis-MSB(20 mg/l) | β | 3.1 | 12.4 | 57.1 | 185.0 | 0.788 | 0.117 | 0.070 | 0.025 |
| PXE/p-TP(3.0 g/l)/ bis-MSB(20 mg/l) | α | 3.1 | 13.4 | 56.2 | 231.6 | 0.588 | 0.180 | 0.157 | 0.075 |

*Table 1. The fast and slow component relative intensities for PC and PXE*

*3.2. Direct measure of the tail-to-total ratio*

The experimental arrangement used in this case is shown in Fig. 5; a small quartz vial is located on the faceplate of a photomultiplier and coupled to it in such a way that the mean number of detected photoelectrons can be properly adjusted. Several replicas of the analog signal produced by the photomultiplier are obtained by a fan-out; they are routed, after being delayed by various amounts, to the inputs of a charge ADC and a constant fraction discriminator, which is used to derive the gate integration signal needed to collect the charge of the pulse.

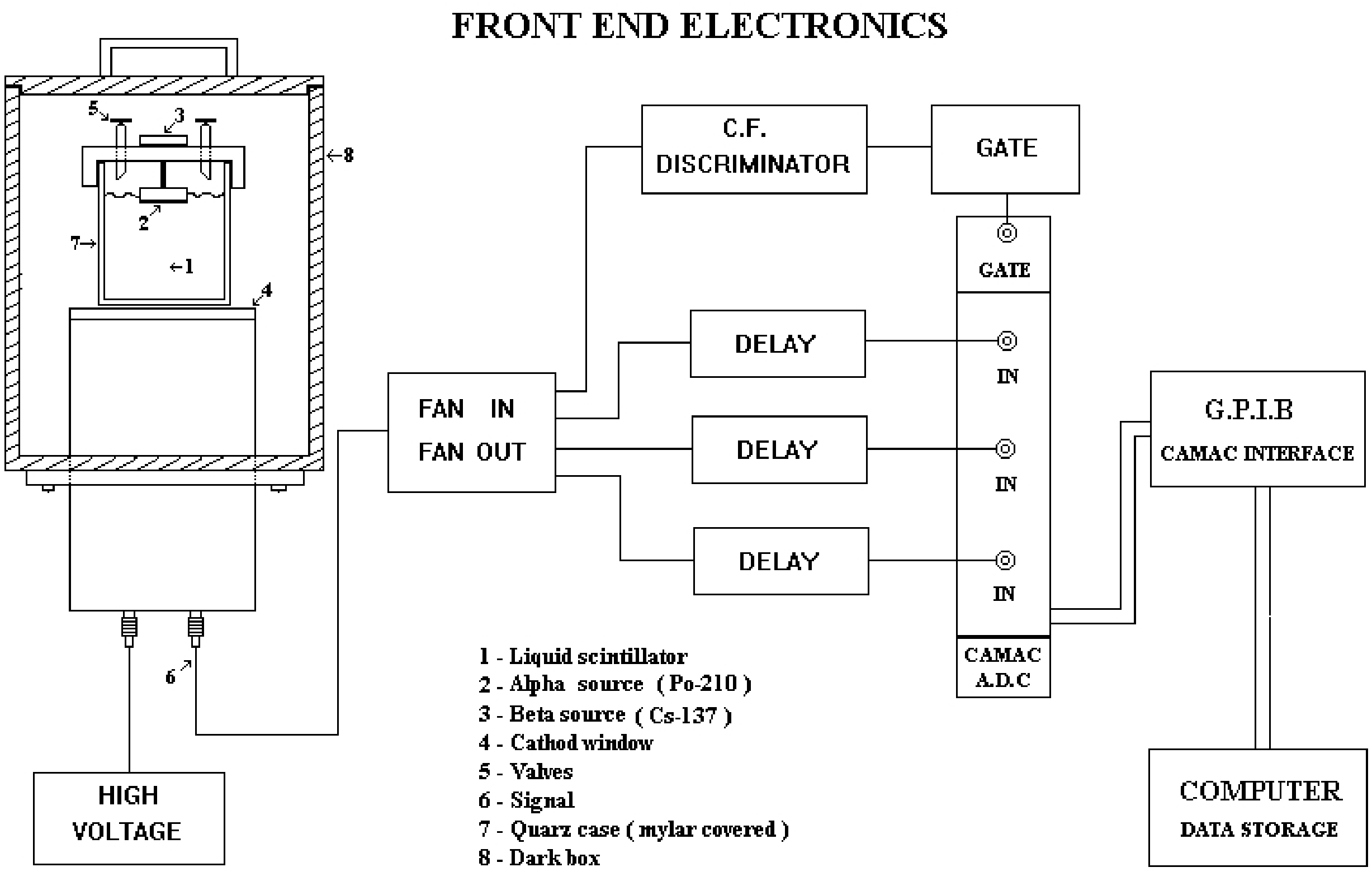

*Fig. 5 - Experimental test set-up for the direct characterization of the PSD properties of liquid scintillator samples through the tail-to-total ratio method.*

With this experimental configuration it is possible to measure the charge for different definitions of the starting point of the tail. In this way we were able to verify that the optimum discrimination is achieved when the beginning of the tail is taken at about 20 ns after the start-up of the pulse.

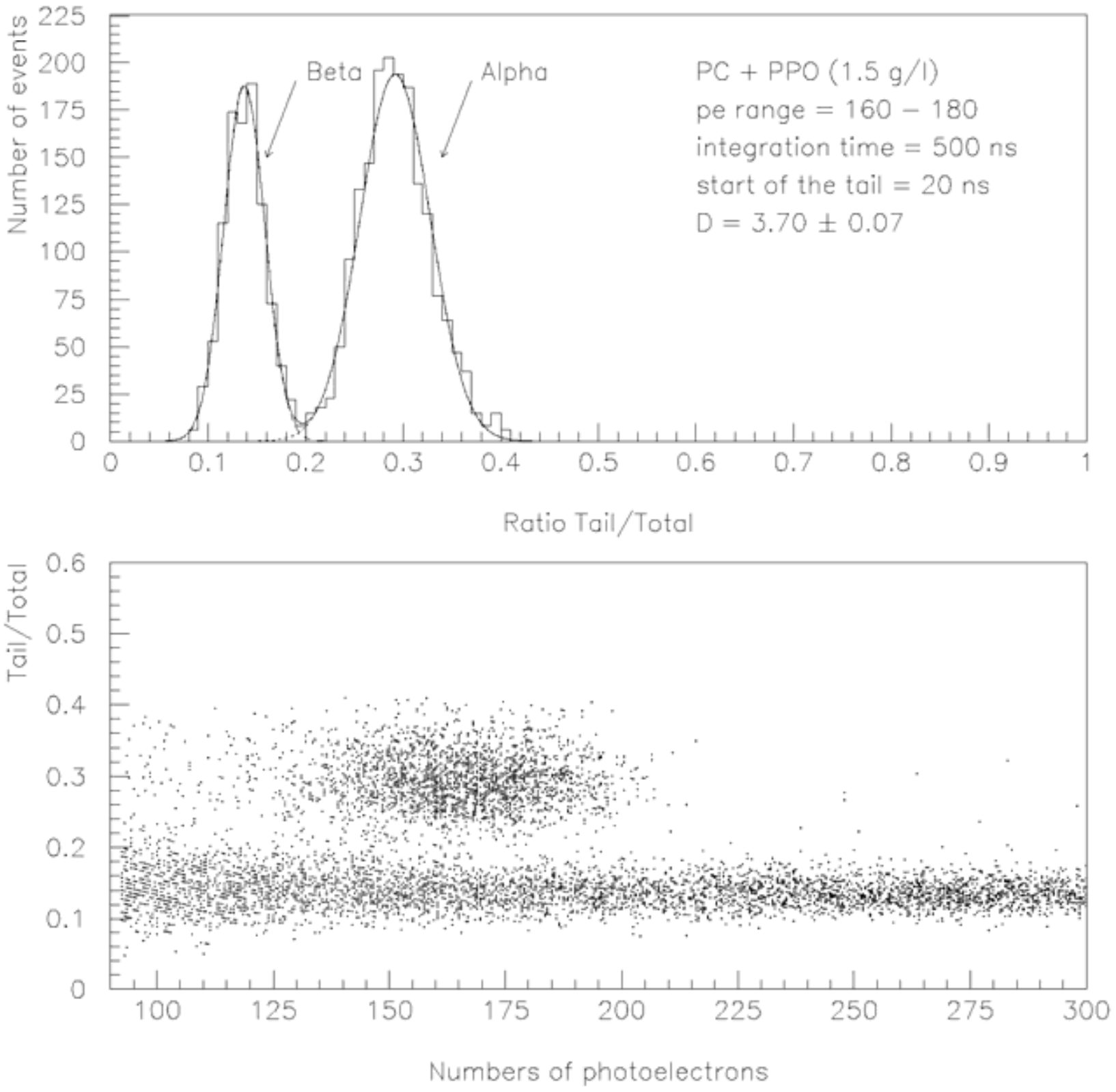

*Fig. 6 - Results of the alpha/beta measurements on the PC+PPO (1.5 g/l) in the scatter-plot format as a function of the number of photoelectrons, together with the projection along the y axis of the interval from 160 to 180 photoelectrons. The measurements have been carried out for an integration gate of 500 ns and for a tail starting at 20 ns.*

Since the direct measurements of the tail-to-total ratio were mainly focused on giving an independent confirmation of the validity of the results provided by the method described in the previous section **§3.1**, we restricted them only to the PC based scintillator. The experimental data are plotted as a scatter plot, according to the format shown in Fig. 6, where the tail-to-total ratio is reported as a function of the number of photoelectrons, together with a projection on the y axis limited to the interval from 160 to 180 photoelectrons. The specific measurement reported in the figure has been carried out with an overall integration gate of 500 ns and the optimum tail starting time of 20 ns.

Both the scatter plot and its projection indicate that the PSD properties of the PC based mixture, in the case of a small size specimen, are excellent, thus further confirming in a completely independent way the indications obtained with the previous methodology.

### *3.3. Quenching*

The complete characterization of the scintillators under study at the level of laboratory measurements in term of α and β responses also required an experimental evaluation of the quenching factor. The light quenching for heavily ionizing particles, i.e., the reduction in the overall light output as a function of increasing values of dE/dx, is connected to the same molecular mechanism responsible for the modification of the pulse shape. In our case of interest, the net effect is that alpha particles of energy $E_\alpha$ produce the same amount of light as beta particles of energy $E_\alpha/Q$, where Q is the so-called quenching factor. While this factor is typically in the range of 10, its precise value depends upon the solvent and the solutes forming the scintillator. Furthermore, even for a given scintillation cocktail of fixed composition, it is also dependent upon the energy of the interacting alpha particles.

For this reason, we investigated the quenching behavior of both the scintillators under consideration through alpha excitations at various energies, obtaining the results which are presented in Table 2. It can be noted that the quenching factor is systematically higher for the PC based scintillator than for the PXE one, and that for both of them the quenching increases as the energy of the alphas decreases. Recalling that the neutrino observation window in Borexino will span the energy interval from 250 to 800 keV, and considering that the alpha particles used in the measurements span the energy range of most of the natural radioactivity alpha background expected in the detector, it is clear from the table that the apparent energies of most alpha events in Borexino will fall well within the neutrino window, thus explaining the need for an effective and powerful PSD approach to selectively remove their contributions to the detected signal throughout the solar neutrino runs.

| Element | α-energy (keV) | Measured energy in PXE (keV) | Quenching factor in PXE | Measured energy in PC (keV) | Quenching factor in PC |
|---|---|---|---|---|---|
| $^{210}$Po | 5300 | 490±10 | 10.8±0.2 | 395±10 | 13.4±0.3 |
| $^{222}$Rn | 5490 | 534±10 | 10.3±0.2 | 410±6 | 13.3±0.2 |
| $^{218}$Po | 6000 | 624±10 | 9.6±0.2 | 483±6 | 12.4±0.2 |
| $^{214}$Po | 7690 | 950±12 | 8.1±0.1 | 751±7 | 10.2±0.1 |

*Table 2. Quenching factor measured for different alpha decays belonging to the $^{238}$U chain*

## 4. PSD hardware implementations in CTF

Both the PSD methods described previously have been implemented in the CTF acquisition electronics. The discrimination processing is applied to the global scintillation pulse, which is obtained by the analog sum of the anode signals of all the phototubes observing the vessel containing the liquid scintillator.

### *4.1. Tail-to-total ratio implementation*

For the tail-to-total (or charge) method, the processing of the global scintillation waveform is carried out via standard charge ADC's and timing units. Essentially, upon the occurrence of a multiplicity trigger signalling a true event, two integration gates are opened, one encompassing the entire signal and providing the denominator of the tail-to-tatal ratio, and the other integrating only the tail.

Since actually in the laboratory studies it was highlighted that the final discrimination performance depends also on the assumed start time of the tail, three delayed integration gates were included in the acquisition electronics, positioned respectively at time offsets of 16, 32, and 48 ns from the start of the pulse. These delays have been imposed through a proper combination of analog delay lines and digital delay circuitries. The digitally integrated charges in the tails provide the numerators of the tail-to-total separation parameters. Experimentally, we found that the best performances in the CTF detector are ensured by the 48 ns tail, and thus all the results reported in the later paragraphs pertain to this case.

It should be noted that the optimum tail offset time in the laboratory measurements was, instead, equal to 20 ns. The discrepancy is likely to result from the modification and distortion effects impacting the tails of the scintillation pulses, described in the next **§5**.

### *4.2. Characteristics of the sampling board (DPSA) optimised for alpha/beta discrimination*

Contrary to the previous standard method, the implementation of the optimum Gatti's filter requires the development of specific circuitries, not readily commercially available.

Hence, for specific use in CTF, we have developed a specialized optimised sampling board, whose main feature are reviewed here.

Preliminarily, it has to be noted that in developing the sampling board, we adopted the modified Gatti's prescription proposed by Jordanov and Knoll [9]. The basic elements of this approach are the preliminary analog integration of the original photomultiplier current pulse and, subsequently, its digitization via a high-resolution 10-bit flash ADC. The combined effect of the pre-integration of the pulse and of the high-resolution ADC is to provide adequate dynamics to achieve the desired sensitivity and accuracy in measuring the pulse for its whole duration, hence providing the basis for the successful application of Gatti's method.

The schematic block diagram of the board is illustrated in Fig. 7. The signal coming from the phototube, after a first amplification stage, goes to the input of a gated integrator, whose integration signal is internally generated by a timing unit driven by an externally produced trigger command, coming from the overall trigger logic of the DAQ system hosting the board. The output analog signal of the integrator is then sampled by two 10-bit, 60 MHz flash ADCs (Analog Devices AD 9020) operating in interleave mode, thus providing an overall sampling frequency of 120 MHz. During the conversion cycle, the flash ADC outputs are stored in two 12 x 256 bit memories.

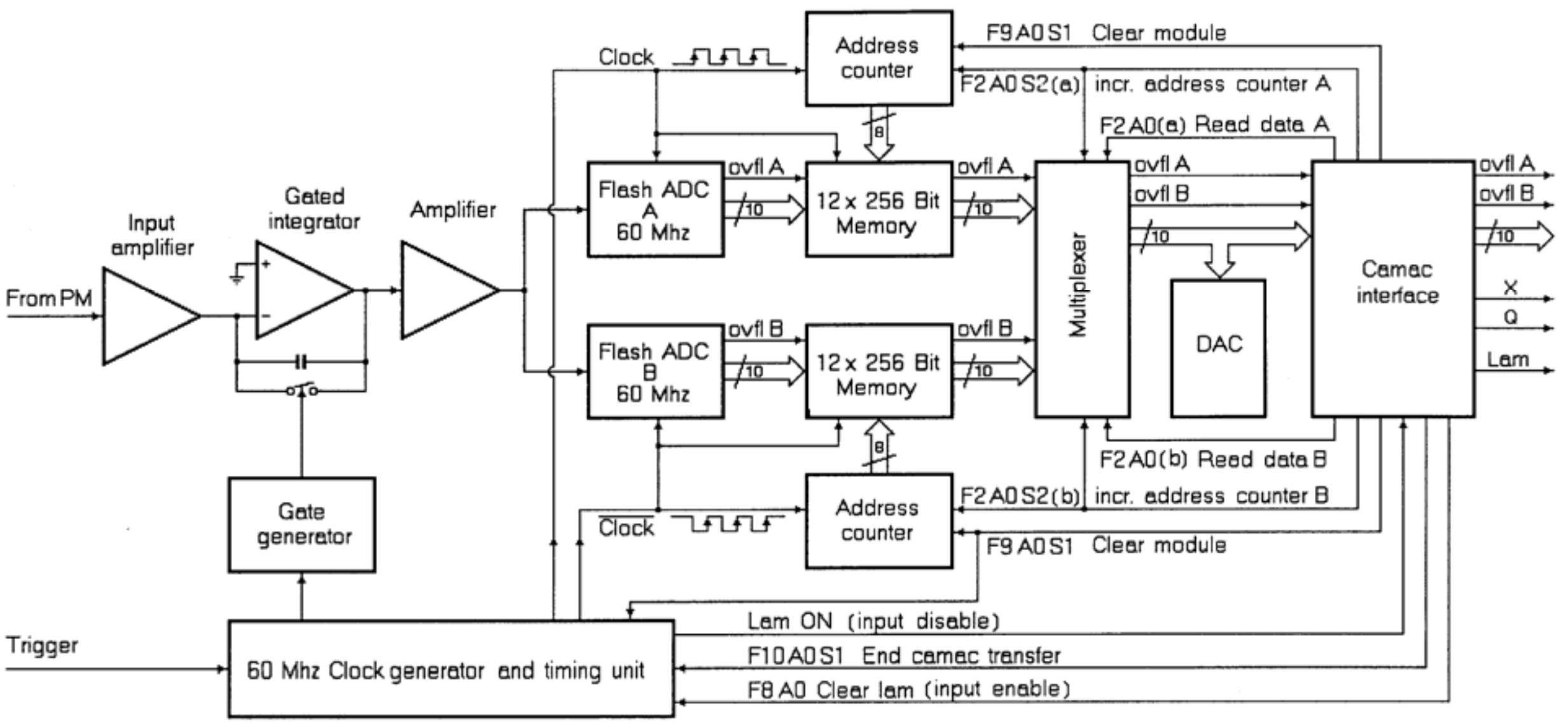

*Fig. 7 - Schematic block diagram of the sampling board (DPSA) optimized for the implementation of Gatti’s method.*

The board was built to the CAMAC standard to ensure compatibility with the existing CTF DAQ system.

The board operates according to three main cycles:

a) The first cycle is devoted to the integration of the input signal, followed by the sampling performed by the two flash ADC's and the recording of the converted data. During this cycle the trigger input is inhibited, as is the CAMAC read out.

b) The second cycle is for monitoring and adjustment, allowing real-time checks of the overall operation of the board. The data from the last conversion are read from the memories and sent, via a multiplexer, to a DAC, whose output can be directly inspected with an oscilloscope.

c) The third cycle is simply that of CAMAC read out.

The three cycles are characterized by different clock frequencies: the first period has the characteristic 120 MHz frequency needed for the interleaving operation of the two ADC's, while the second monitor cycle is operated at 10 MHz. The board is thus equipped with two subsections which deliver both of the required clock frequencies. The 1 MHz operation of the third cycle, the CAMAC one, is dictated by the CAMAC bus itself.

Given the low rates expected in our application, the capability of the board to cope with high count rates is not an issue; regardless, the board in the test experimental set-up proved to be capable of detection rates of 1 kHz, and this limit is due only to limitations of the CAMAC architecture.

The last architectural feature to mention is the possibility to toggle the operating condition of the board between remote CAMAC mode and local mode, a very useful capability while performing any preliminary tuning and checking operation.

In Fig. 8 the analog portion of the board is shown in detail. The first stage provides the amplification of the signal. The second stage is the gated integrator, whose output is the signal integrated over a period defined by the duration of the relevant driving command from the timing unit. The operational amplifiers used in these stages have adequate bandwidth for the signals being processed.

The signal integration command has two parts: the first is the actual signal command to integrate the pulse, and the second is the command for integration of the baseline.

The durations of the integration time, of the reset time, and of the baseline integration time are variable, and can be adjusted via internal regulators.  Any combination is possible, within the limit that the sum of their respective values cannot exceed 4 μs. It should be added that by properly adjusting the trigger and signal relative delay, the baseline integration is actually performed twice, once before and once after the actual pulse, thus leading to an unambiguous and precise assessment of the "zero" value.

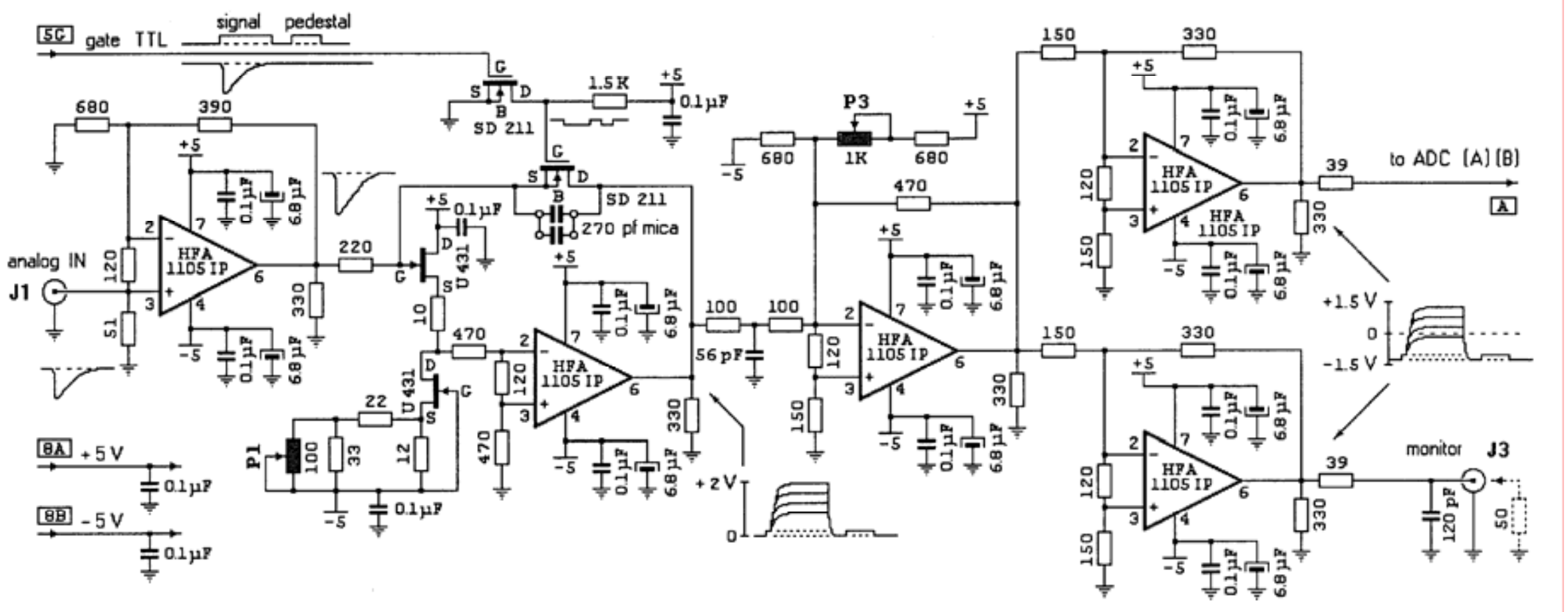

*Fig. 8 - Analog portion of the DPSA sampling board.*

The potentiometer P1 shown in Fig. 8 has the purpose of exactly balancing the integrator output. When this condition is achieved, the integrated output, with no signal at the input, is perfectly flat. The other potentiometer P3 allows one to define the dynamic range of the integrated signal presented at the input of the ADC's, which has to be symmetric with respect to the zero value and not exceeding, for linear operation, an absolute value of 1.5 V.

The flexibility and versatility of the board is also ensured by the presence of an internal test signal that emulates two different pulse shapes, and which can be exploited to facilitate the overall system adjustment.

Examples of waveforms detected with the board in a small scale laboratory set-up, prior to its use in CTF, are shown in Fig. 9. More details on the features of the board can be found in [10].

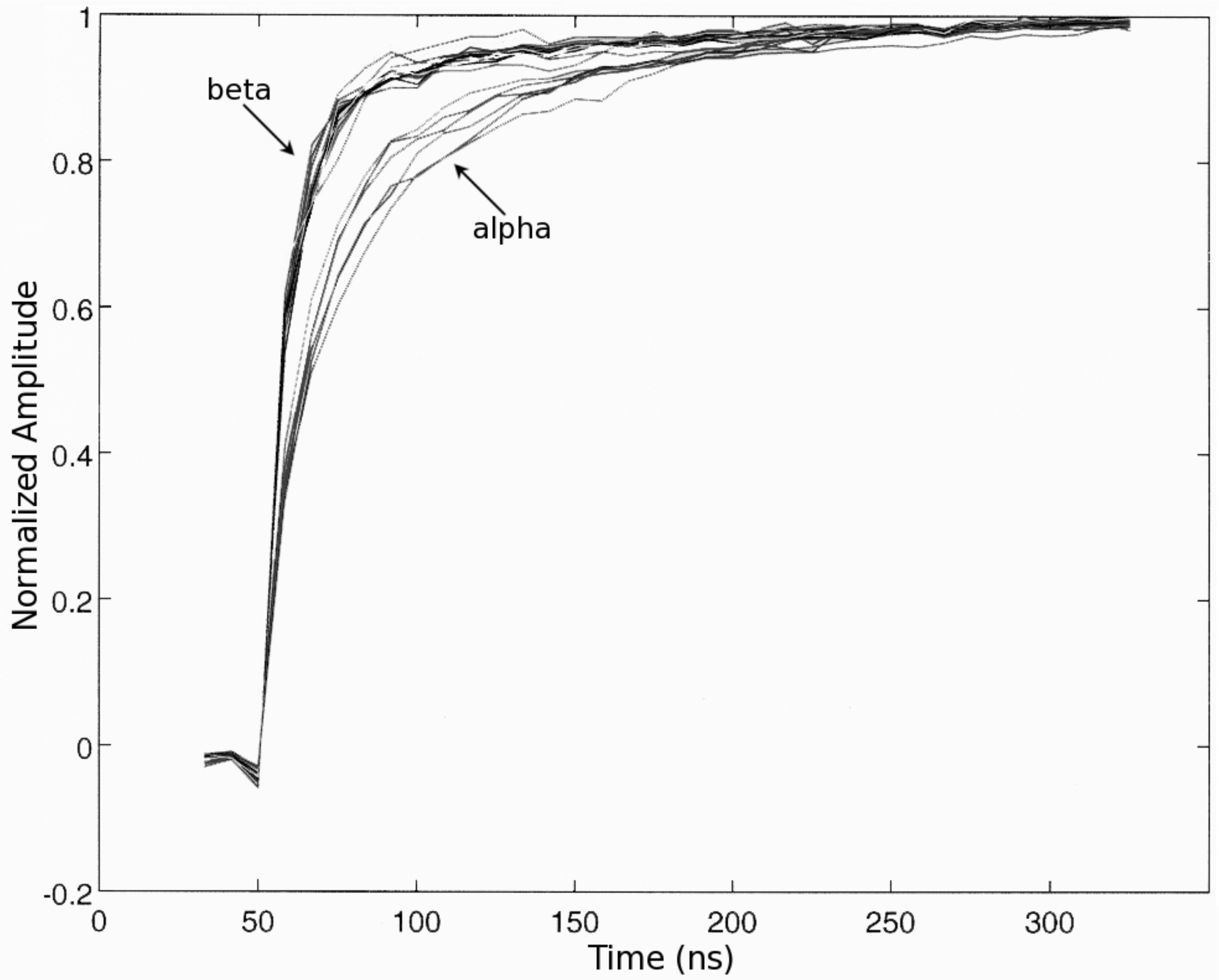

*Fig. 9 - Examples of waveforms, produced by a scintillator under $\alpha$ and $\beta$ irradiation, integrated and acquired by the sampling board.*

## 5. Light propagation effects across the CTF volume

While passing from the small scale laboratory samples to a massive volume of scintillator, such as CTF, the features of the scintillation light are altered by effects which, though undetectable in limited size specimens, manifest clearly as photons travel across an extended volume.

There are essentially two main photon effects, subjects of thorough studies reported in [11] and [12]: the Rayleigh scattering and the absorption re-emission process. It should be pointed out that the latter phenomenon is due to the overlap of the absorption and emission spectra of the scintillator, so that a photon originally produced at the primary interaction site can undergo one or more steps of self-absorption followed by re-emission.

Both propagation effects produce a modification, specifically a lengthening, of the overall light pulse, as detected by the ensemble of observing phototubes, with respect to the ideal shape acquired in the laboratory tests. In particular, since the tails of the pulses represent the portions of the signals more affected by the large scale propagation mechanism, it is reasonable to expect a worsening of the PSD characteristics of the scintillator in an extended volume setup, compared to those exhibited by the same scintillator material placed in a limited size counter.

On top of this, one cannot ignore the lengthening effect of the scintillation pulse simply due to the geometrical effect of the differences between path lengths of the various photons from a scintillation event to the observing phototubes.

In order to unravel the main characteristics of these effects, the CTF time response has been studied through the acquisition either of signals induced by sources immersed in the scintillator, or of events originating from some of the intrinsic, small, contaminations of the scintillator.

For example, the study of events produced by a point-like, confined Radon source deployed in the center of the vessel yielded a first pulse time component of decay time ~5 ns, a significant worsening with respect to the 3.5 ns value measured in small samples for the PC based cocktail. Observation of events generated by the low-energy decays of $^{14}C$ present in the scintillator (which in practice acts as a uniformly distributed volumetric source) gave the same result.

It is thus reasonable to expect a similar effect on a much longer scale, regarding the tail of the pulses; however, in CTF it was not possible to study the time evolution of the tail of the scintillation signals with the same accuracy as the laboratory measurements, for two reasons: first, the lower statistics attainable (only sources with very low activity could be deployed, and any intrinsic contamination used for self-calibration also provides low-statistics measurements because of the necessarily low-background characteristics of the detector); and second, the shorter acquisition range of the CTF time-to-digital converters (TDC), limited to about 110 ns (to be compared to the μs scale of the laboratory measurements, see Fig. 4). Despite this, measurements of the scintillation pulse shapes with the CTF electronics have been carried out, with results that will be shown in the next section **§6.1.**

## 6. Experimental evaluation of the α/β discrimination capabilities in the volume of the CTF detector

The results of the evaluation of the CTF α/β discrimination capabilities reported in this paragraph have been obtained using data acquired, respectively, in July - August 2000 with the PXE-based scintillator, and in December 2001 - January 2002 with the PC+PPO scintillator.

The first part of the analysis strategy is based on the identification of clean alpha and beta samples, through the so-called delayed coincidence technique. Afterwards, once the reliability of the two PSD methods under consideration has been demonstrated on these known alpha and beta events, the two techniques are applied to samples of events of unknown composition.

### *6.1. Analysis of α and β reference pulses*

For the reference analysis, we perform a preliminary study of data which are independently known to be alpha or beta signals. For this purpose we exploited the data taking periods during which the detector was affected by a sizeable, though limited, Radon contamination. Indeed, among the radioactive products which follow the decay of the parent Radon element, there is the distinctive correlated sequence $^{214}$Bi-$^{214}$Po, characterized by a mean time delay of 237 μs, which can be easily identified because of the short time difference between the first and second decay in the sequence. Since the precursor, the $^{214}$Bi (β+γ emitter with end-point 3.23 MeV), induces β-like pulses in the scintillator, while the successor $^{214}$Po is an alpha emitting nuclide ($E_{\alpha}$=7.668 MeV), such a sequence provides an effective way, independent from any PSD technique, to build two clean reference samples of alpha and beta signals.

The specific selection criteria for this delayed coincidence identification are that the coincidence time between consecutive events must fall in a window between 2 and 710 μs (3 times the mean life of $^{214}$Po) and that the energy of the second event must be found in a range of 3σ around the expected mean value of the $^{214}$Po apparent energy (950 keV in PXE and 751 keV in PC, taking into account the quenching factors of table 2). The lower threshold of 2 μs in the time coincidence requirement is adopted to reject the short $^{212}$Bi-$^{212}$Po coincidence in the $^{232}$Th chain, characterized by a mean decay time τ=433 ns. The overall efficiency of these cuts is 95%, while the contamination in the event sample of unwanted signals due to accidental coincidences is negligible (<1%). In the plots of Fig. 10 the energy distribution of the first and second events are shown, while in Fig. 11 the time difference between the two events is plotted and fitted to confirm the correct time behaviour of the coincidence.

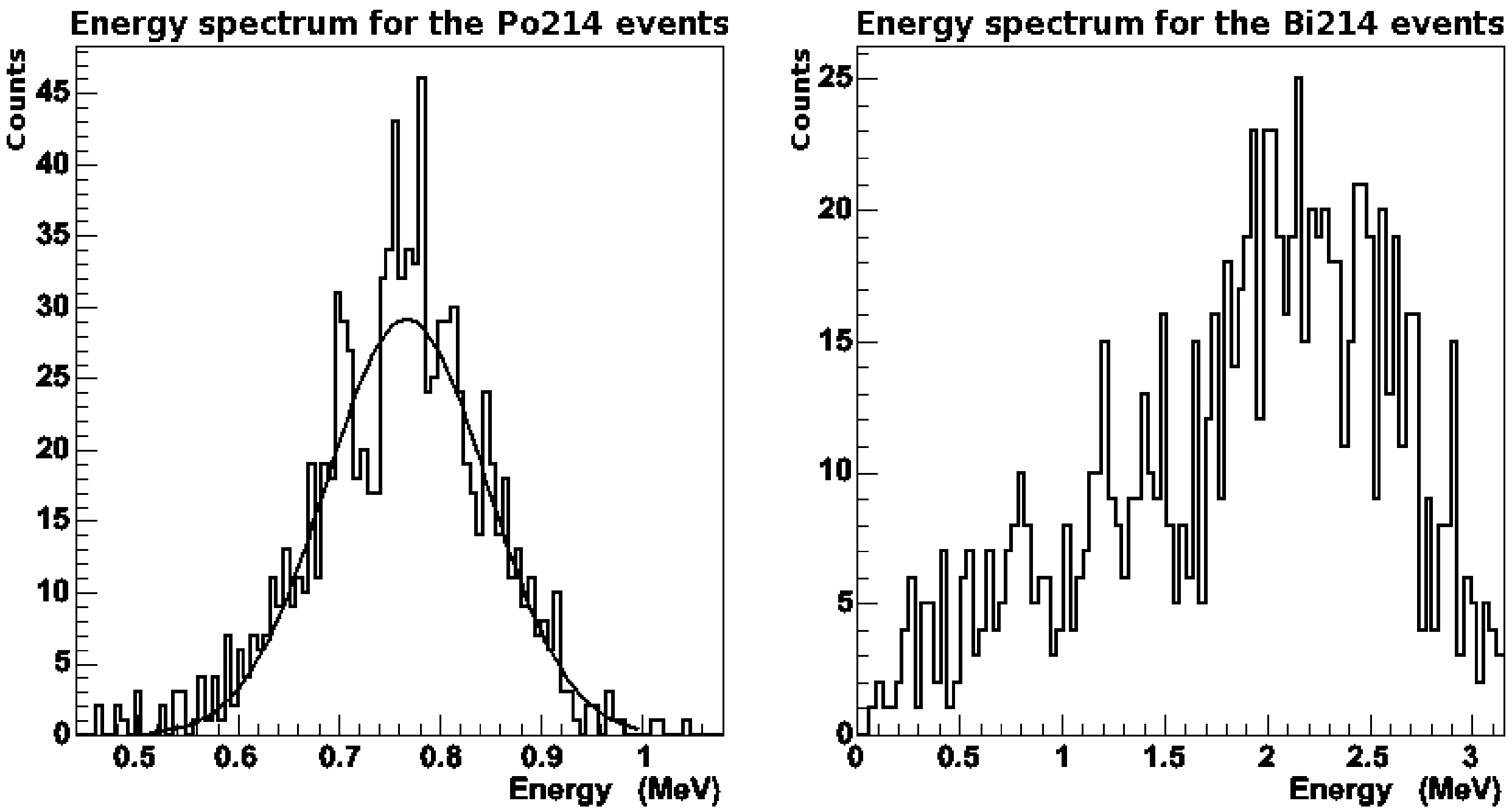

*Fig. 10 - Energy spectra of $^{214}$Po (left) and $^{214}$Bi (right) events detected in PXE, selected via the delayed coincidence method.*

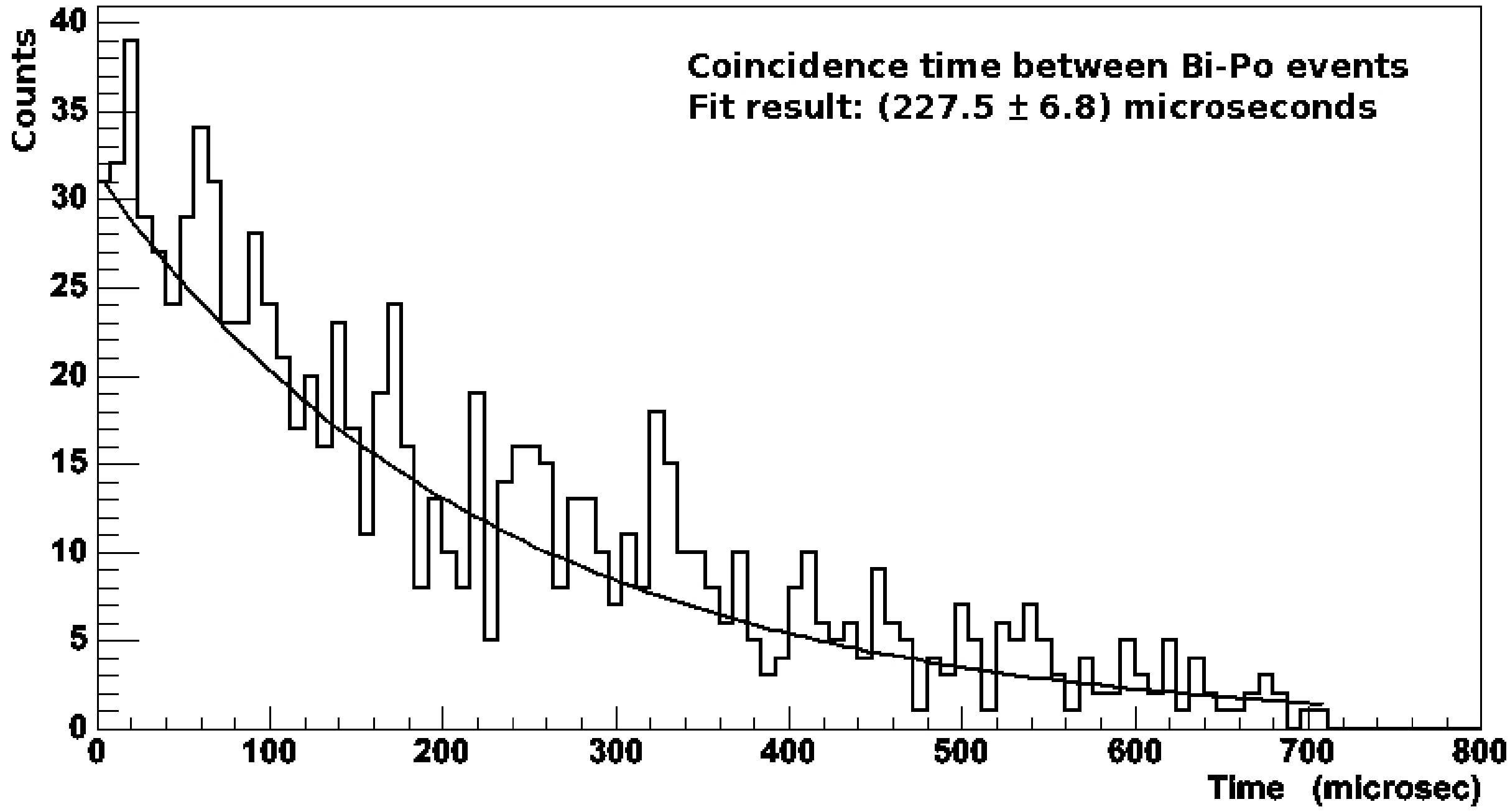

*Fig. 11 - Measured time difference between pairs of $^{214}$Bi and $^{214}$Po events; $\tau(^{214}Po)$ =236.6 $\mu$s.*

Actually, in our procedure we used two different $\alpha$ and $\beta$ reference sub-samples (both selected according to the above methodology): the first sub-sample was used to tune the hardware modules used both for the tail-to-total and the Gatti's filter methods, providing in particular for the latter the reference pulse shapes required to adjust the processing of the output of the associated sampling board; afterwards, the second independent sub-sample was used to quantify the $\alpha/\beta$ discrimination efficiency of both Gatti's technique and the tail-to-total method in the range of the $^{214}$Po energy.

Before proceeding to the core of the $\alpha/\beta$ analysis, we exploited the reference pulses also to investigate the pulse shape of the signals, thus extending to CTF the same kind of study reported in **§3.1**, obviously within the limit of the statistics and accuracy permitted by the intrinsic limitations of CTF, as already explained in the discussion in **§5.**

The output of this pulse inspection is twofold and is reported in Fig. 12 and 13; the former shows the pulse shape of the alpha and beta reference scintillation signals acquired through the "standard" CTF electronics over a 110 ns interval, from which we can infer that the basic difference in shape between the two types of events is preserved even in a "large volume" condition such as that of CTF. The latter, instead, displays the time evolution of the integrated alpha and beta average reference pulses over a longer time range of almost 400 ns, as obtained from the pre-processing section of the sampling board. Thus, the signals originated by the board pre-elaboration, also maintain the difference of the two type of events, and, remarkably, over a significantly longer period with respect to that explored by the standard TDC's, which is surely beneficial for the expected PSD performances.

It can be noted that, contrary to the monotonic increasing behaviour of the laboratory integrated waveforms in Figure 9, the CTF integrated signals reach a maximum and then start to decrease. The origin of this discrepancy is the different AC coupling of the anode output circuit adopted in CTF with respect to the laboratory set-up. In CTF, indeed, we employed the AC output developed for the optimum coupling to the Borexino front-end, which produces this peculiar behaviour of the output integrated pulses.

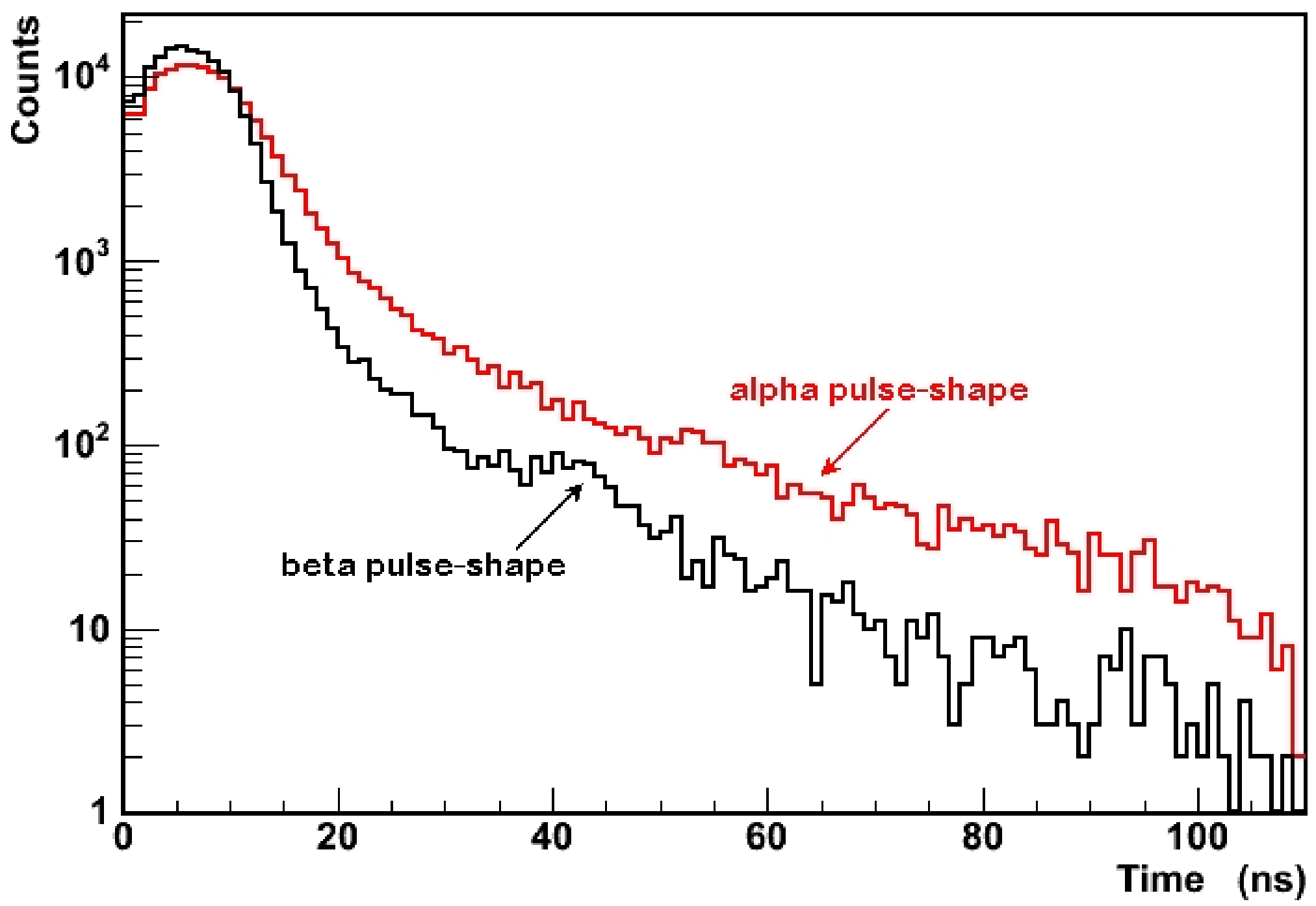

*Fig. 12 - Examples of α (red curve) and β (black curve) pulse shapes, as measured in CTF through the "standard" electronics.*

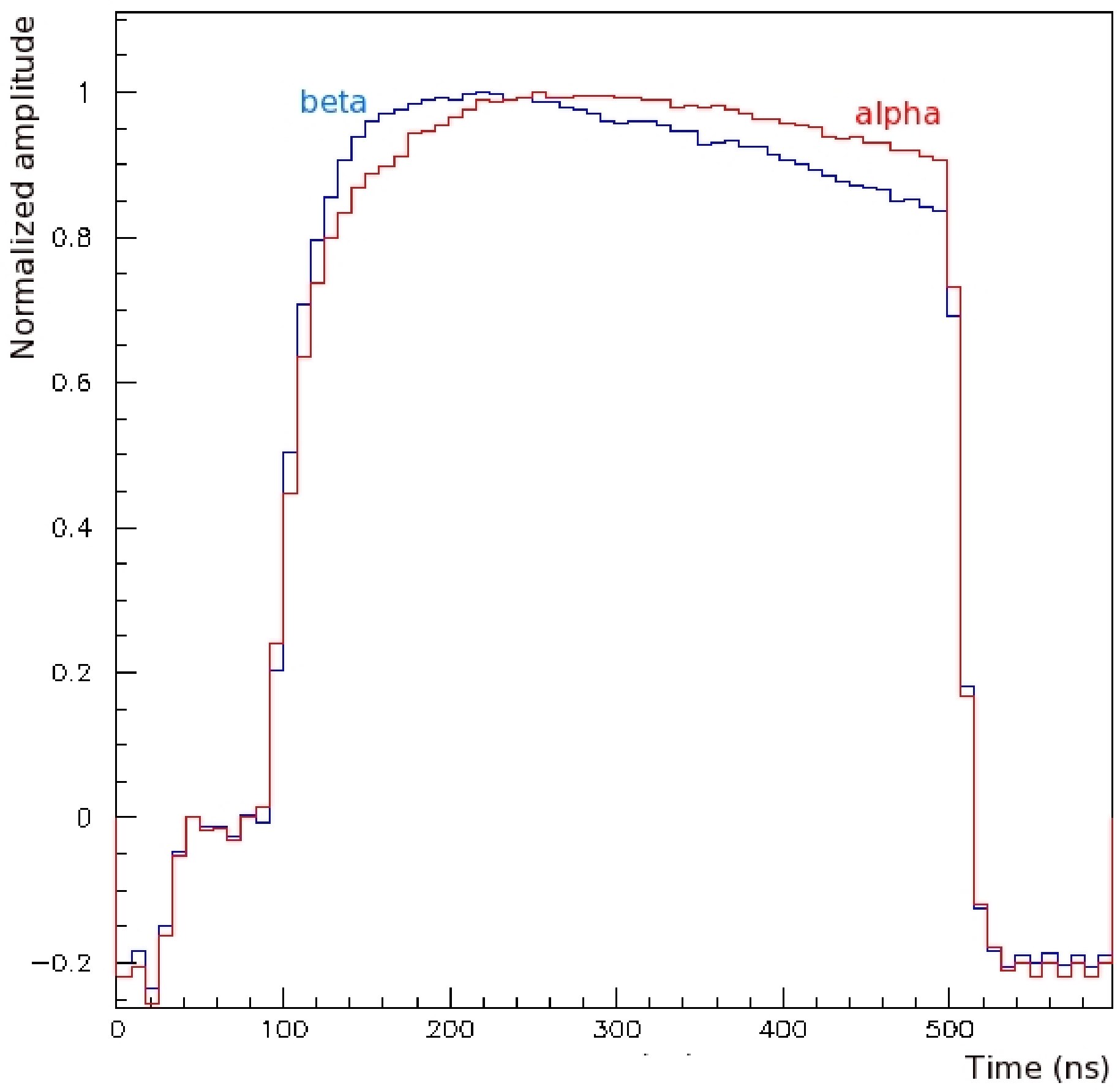

*Fig. 13 - Examples of α (red curve) and β (blue curve) pulse shapes, as acquired in CTF through the DPSA board.*

After the previous pulse shape investigation, we went ahead and used the first sub-sample of reference pulses to compute the optimum filter weights $P_i$, (eq. (3)) according to the prescription of **§2.2**, and then to check the proper adjustment of both the Gatti's and tail-to-total hardware. The results of this PSD "reference" adjustment and analysis in the PC based scintillator are reported in Fig. 14, where we show, respectively, the distribution of the tail-to-total ratio and of the Gatti parameter for the selected reference alpha and beta signals in the $^{214}$Po energy range. From the figure we can infer that both the PSD hardware sections are well tuned, producing in all cases clean Gaussian distributions of the separation parameters. Obviously, in each display the amount of overlap between the alpha and beta distributions is a qualitative measure of the discrimination power: ideally for 100% separation there should be zero overlap, which is not the case in the figure.

It can also be noted that the amount of overlap is less in the Gatti's case, as expected because of the absolute optimality of this method.

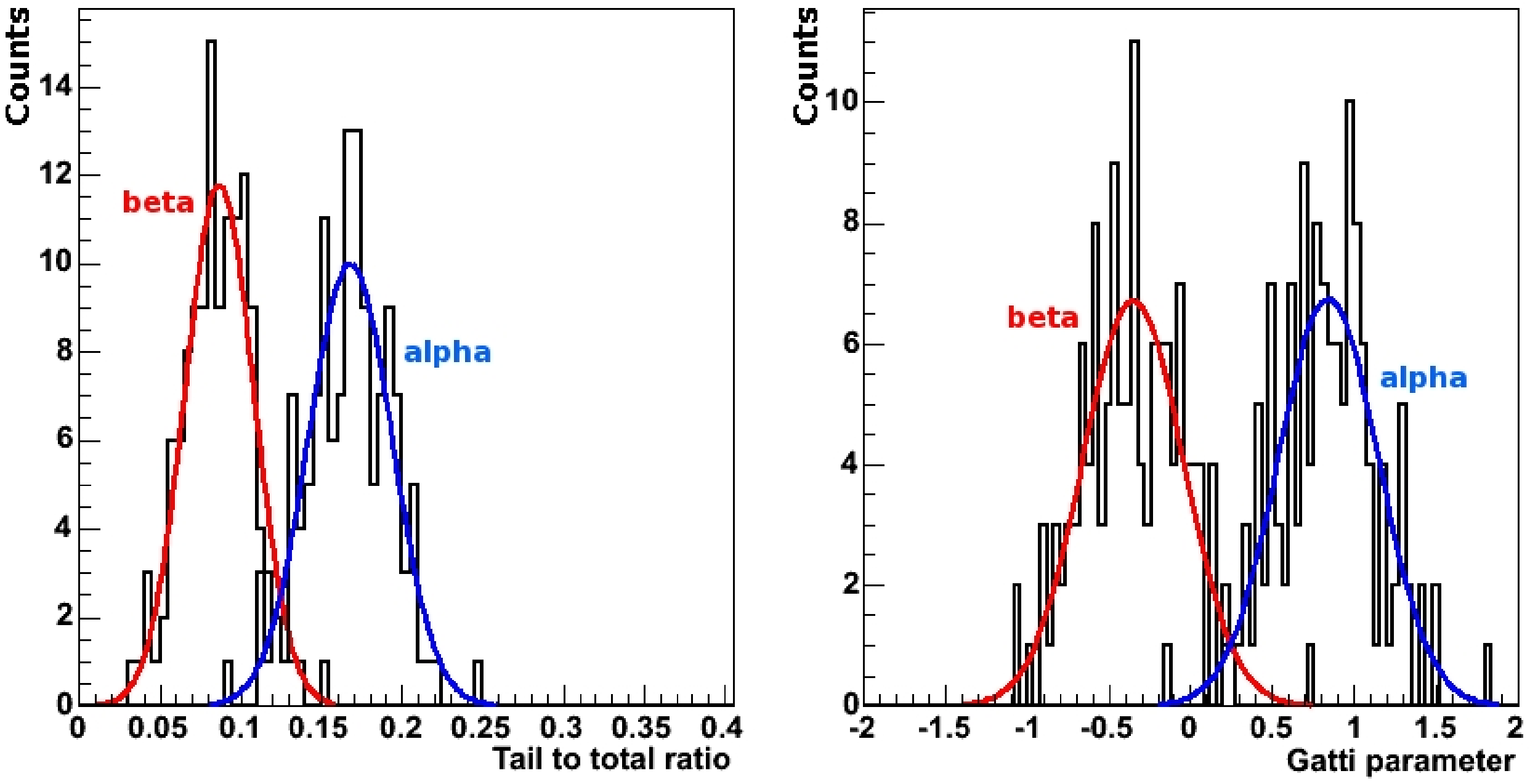

*Fig. 14 - Distributions of the α/β discrimination parameters, tail to total (left) and Gatti (right), for $^{214}$Bi-$^{214}$Po events detected in PC+PPO (β distributions are fitted in red and α distributions in blue).*

### *6.2. Evaluation of the discrimination performances using an independent sub-set of the reference pulses*

Once the preliminary reference investigation was completed, the second sub-sample selected via the $^{214}$Bi-$^{214}$Po coincidence technique was employed for a quantitative assessment of the discrimination capability obtained in the CTF detector, at the energy of the $^{214}$Po decay. Specifically, this study has been done as a function of the estimated positions of the scintillation pulses within the scintillator vessel, as derived from the recorded photon arrival times on the phototubes. In particular, in Fig. 15 the tail-to-total discrimination parameter is plotted in different spatial regions (this time for a graphic example we used the PXE data): in the first plot, all $^{214}$Bi-$^{214}$Po coincidences throughout the entire vessel are considered; in the second plot a radial cut of 90 cm is applied to the events, to avoid the effect of photon reflection and refraction at the boundary between the scintillator and the water; in the third plot a radial cut of 60 cm is applied, to minimize the dispersion on photon arrival times due to scattering, absorption and re-emission, and geometrical effects. It is intuitive that while passing from the entire vessel to its central core the overall discrimination properties should improve: indeed, all the effects mentioned above, which are more pronounced in the external shells of the vessel, tend to distort the pulse shape with a lengthening of the tails that impacts the capability to correctly distinguish the pulses of different origin.

Indeed, this behaviour is visible in Fig. 15, where the degree of overlap between the alpha and beta distributions clearly diminishes as a function of the tightness of the radial cut.

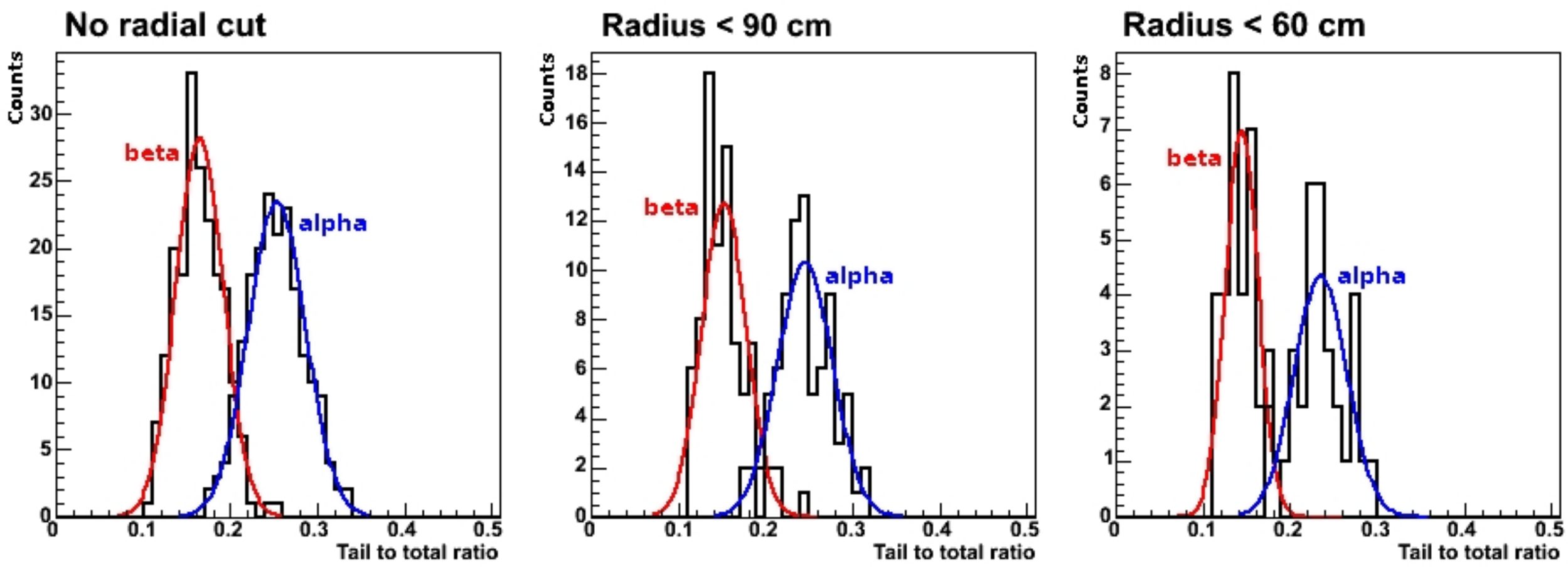

*Fig. 15 - Behaviour of the tail-to-total discrimination parameter, for $^{214}$Bi-$^{214}$Po samples selected with different radial cuts (in PXE). $\beta$ distributions are fitted in red and $\alpha$ distributions in blue.*

We have performed the kind of analysis reported in Fig. 15 systematically for the two scintillators and the two methods considered in this work, obtaining the results summarized in Table 3. Specifically, in the table we have chosen to report quantitatively the degree of overlap of the alpha and beta curves (which is the measure of the PSD discrimination power) by listing the alpha mis-identification probability for a fixed beta identification probability of 98%. The general results are very good: the probability to mis-identify an alpha particle is always lower than 8% (corresponding to an efficiency in $\alpha$-type background rejection of 92%), in the case of PC+PPO scintillator. In particular, the optimum Gatti's technique gives generally better performances than the tail-to-total method ($\alpha$-rejection efficiency is at least 97%), and the PC+PPO scintillator shows significantly better performance than the PXE-based one. This last comparison was possible only for the traditional tail-to-total technique, because for the PXE sub-sample used in the present analysis, the DPSA board was not yet available.

| Scintillating mixture | $^{214}$Po energy range (keV) | tail/tot (no cut) | tail/tot (90 cm) | tail/tot (60 cm) | Gatti (no cut) | Gatti (90 cm) | Gatti (60 cm) |
|---|---|---|---|---|---|---|---|
| PC/ PPO(1.5 g/l) | 550 – 950 | 7.6% | 5.4% | 2.3% | 3.3% | 1.3% | 0.37% |
| PXE/p-TP(3.0 g/l)/ bis-MSB(20 mg/l) | 700 – 1200 | 16% | 10% | 4.4% | N.A.$^{(*)}$ | N.A. | N.A. |

*Table 3. $\alpha/\beta$ discrimination capabilities in the high energy range, for the two scintillating mixtures, with two discrimination methods and for different radial cuts; results are quoted for a fixed beta identification probability of 98%: the quantity reported in the table is the alpha mis-identification probability.*

*$^{(*)}$Sampling board not yet installed*

### *6.3. Evaluation of the discrimination performances for lower energy alphas*

To conclude the evaluation of the CTF PSD discrimination capability, we performed a replica of the thorough investigation (at the energy of the $^{214}$Po events) reported in Table 3, in a lower energy range. The need for this extension is twofold: on one hand, it is known that the PSD performances depend upon the energy, and it is thus interesting per se to know in this respect the actual behaviour of the scintillators under study; and on the other hand, for the specifics requirements of the Borexino solar neutrino investigation, the major and more dangerous alpha background is due to nuclides emitting alpha particles in the energy range of 5-6 MeV [2].

To perform this investigation, we exploited the occurrence in the scintillator, in the period of the sizable Radon contamination and prior to the final purification process, of signals produced by the three alpha emitters $^{222}$Rn, $^{218}$Po and $^{210}$Po (for their energies and the corresponding quenching factors, see again Table 2).

It must be pointed out that in the range of interest of the low energy alphas, the CTF data are affected also by muon induced events, which can affect the output of the $\alpha/\beta$ discrimination investigation. To remove this latter category of signals, thus ensuring only the selection of proper scintillation events, the CTF muon detector was used, which allows rejection of cosmogenic pulses with an efficiency larger than 99% in the relevant energy region [12]. The energy spectrum of the residual events is shown in Fig. 16, together with the spectrum prior to the muon rejection, in the case of PXE-based scintillator (note that the $^{40}$K peak shown in the figure is not an intrinsic feature of the scintillator, but an external background contribution coming from some of the surrounding materials). The bump located at about 400 keV in the muon-subtracted spectrum is due to the $^{222}$Rn, $^{218}$Po and $^{210}$Po alpha induced signals, superimposed on a continuous spectral contribution of beta events.

As an explicative example, Fig. 17 shows the tail-to-total and Gatti's parameters for these classes of events, in the PXE scintillator. Since the two typologies of events are mixed in an unknown proportion, the experimental histograms of the separation parameters are also mixed together for alpha and beta events. In other words, we do not have now two Gaussian curves describing individually the experimental distributions of the separation parameters, as in the case of the $^{214}$Bi-

$^{214}$Po reference pulses, but only the sum of the two distributions is an experimentally accessible quantity.

In order to unravel the discrimination power at low energy from the cumulative histogram, it is thus necessary to perform a fit to the sum of two Gaussian curves, as shown in Fig. 17. The curves resulting from the fit, indeed, are the best possible estimates of the individual distributions of the separation parameters, leading to a quantitative assessment of the desired PSD performances, following the same criterion of alpha mis-identification for a given beta identification probability applied in the previous Table 3.

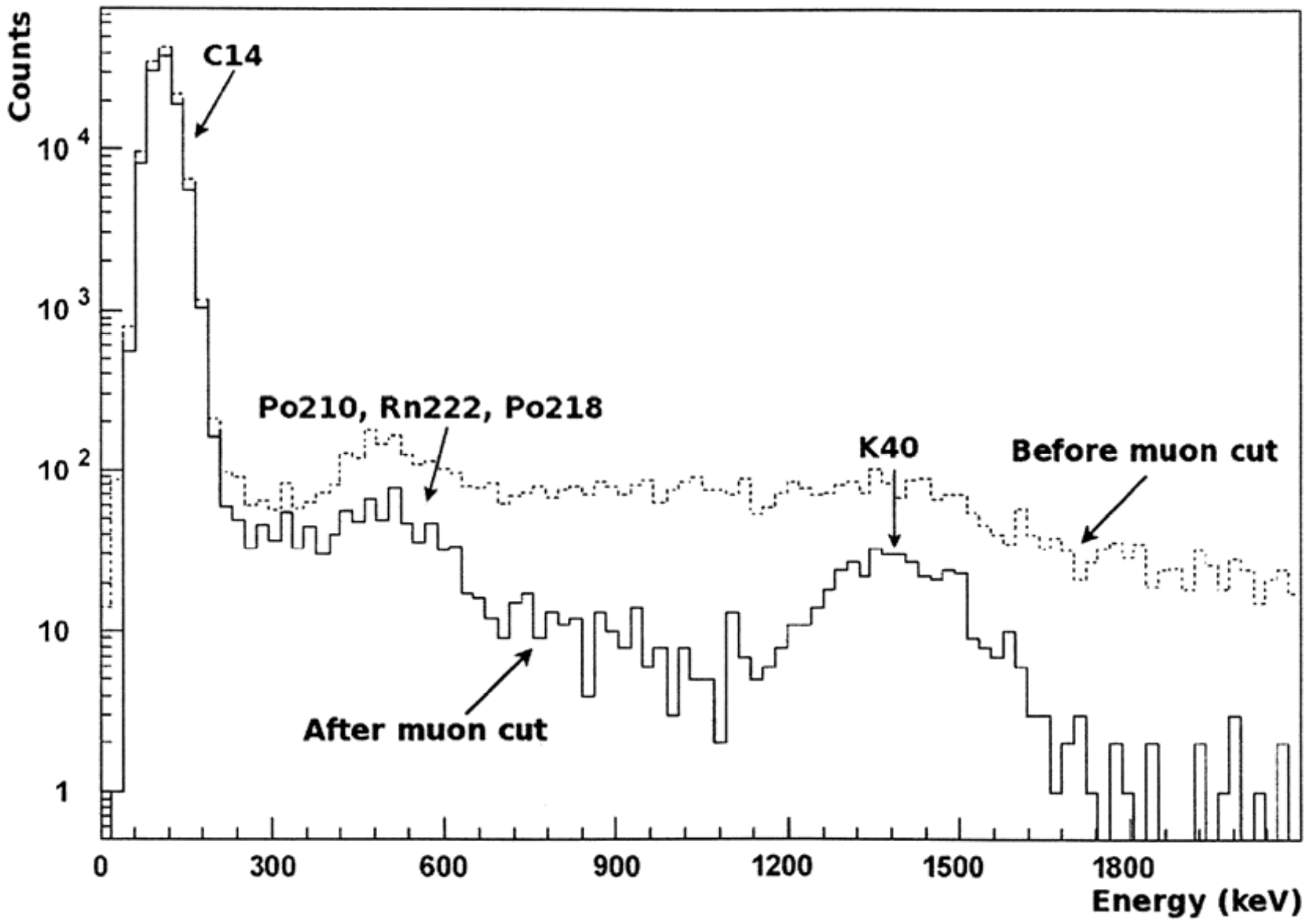

*Fig. 16 - The CTF spectrum acquired with PXE scintillator: all events with r < 90 cm are shown, before muon rejection (dotted line) and after muon rejection (solid line). The bump produced by the low energy alphas is clearly enhanced in the spectrum after the muon rejection.*

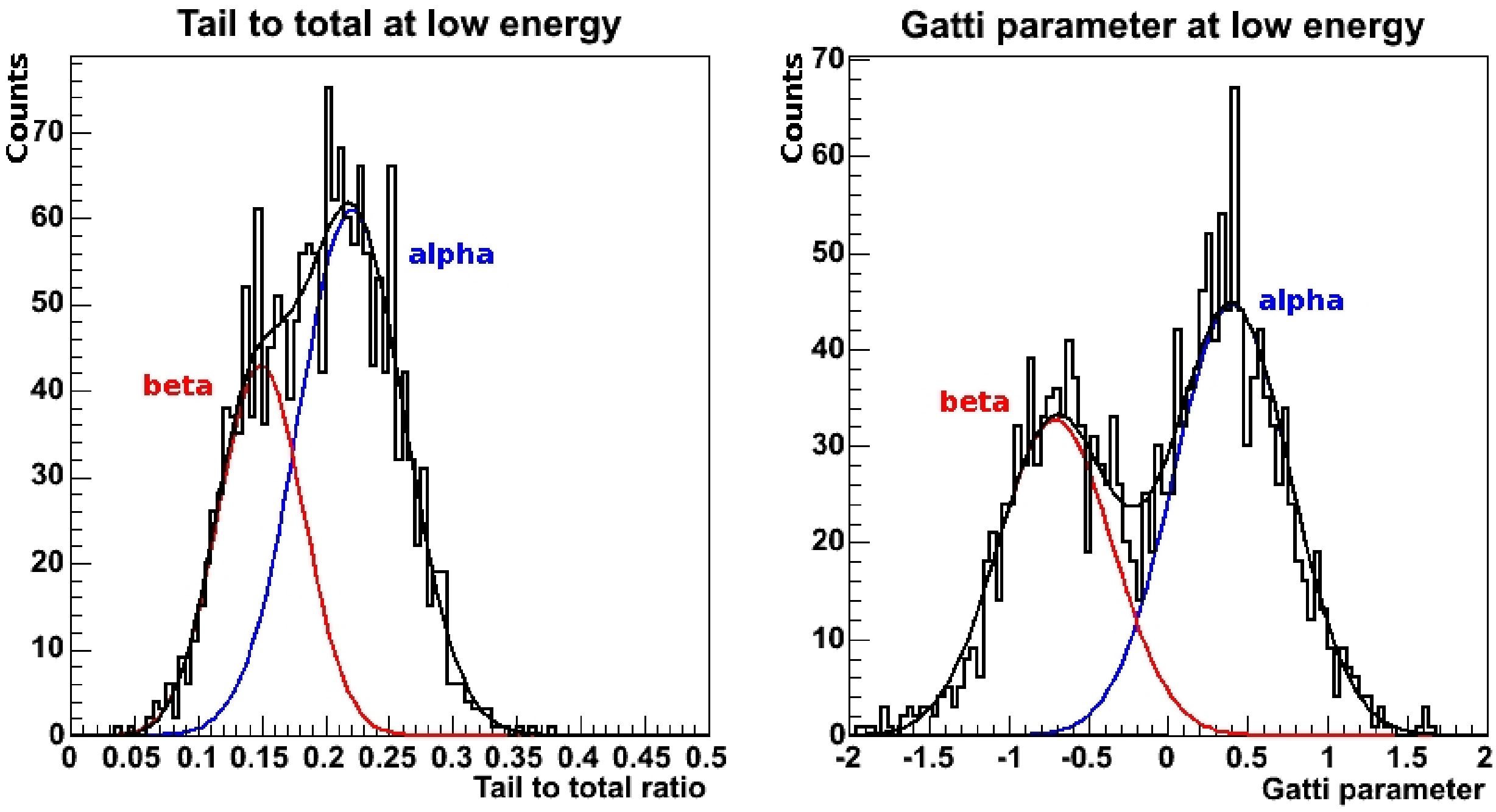

*Fig. 17 - Performances of the $\alpha/\beta$ discrimination parameters in the energy range 200 - 700 keV, in the case of PXE-based scintillator: the tail-to-total parameter (left) and the Gatti parameter (right). $\beta$ distributions are fitted in red and $\alpha$ distributions in blue.*

In Table 4 the results of this assessment are reported for both liquid scintillators and the two discrimination parameters, and assuming 95% beta identification probability. This value, more conservative than the 98% assumed above for the higher energy $^{214}$Po, coincides with the Borexino targeted design value. From the table we can infer the good PSD performances featured by the PC based scintillator with both methods, even though, as expected, Gatti's filter behaves in a better way. In this respect it is quite remarkable, in view of the application in Borexino, that deep in the center of the vessel, Gatti's method ensures an inefficiency in the tagging of the low energy alpha events limited to 1.5%.

On the other hand, the Table confirms that the PXE scintillator features intrinsically worse $\alpha/\beta$ capabilities. For example, in the case of the tail-to-total ratio, the analysis provided a meaningful result only for the 60 cm radial cut, even in this case giving a very poor alpha tagging inefficiency value, as high as 22%. For this PXE analysis, Gatti's method was also possible (which was not the case in the reference pulse analysis of Table 3), and thus from the Table it can be noted how it helps to drastically improve the PXE PSD, which now in the 60-cm core of the vessel reaches the definitively better value of 7.8% (a value nevertheless worse than the corresponding performance of the PC scintillator).

| Scintillating mixture | low energy range (keV) | tail/tot (no cut) | tail/tot (90 cm) | tail/tot (60 cm) | Gatti (no cut) | Gatti (90 cm) | Gatti (60 cm) |
|---|---|---|---|---|---|---|---|
| PC/ PPO(1.5 g/l) | 200 – 550 | 6.5% | 5.7% | 3.9% | 5.3% | 2.5% | 1.5% |
| PXE/p-TP(3.0 g/l)/ bis-MSB(20 mg/l) | 200 – 700 | N.A.(*) | N.A. (*) | 22% | N.A. (*) | 13% | 7.8% |

*Table 4. α/β discrimination capabilities in the low energy range, for the two scintillating mixtures, with two discrimination methods and for different radial cuts; results are quoted for a fixed beta identification probability of 95%: the quantity reported in the table is the alpha mis-identification probability.*

(*)*The two classes could not be separated in this data sample*

## 6. Conclusion

The main outcome of this work is that we have demonstrated the possibility to perform the α/β discrimination in a large volume liquid scintillation detector, such as the Borexino Counting Test Facility. Both of the PSD methods tested, i.e., the optimum Gatti's filter and the very popular charge (or tail-to-total ratio) technique, proved to be effective tools in discriminating particles of different nature, though Gatti's method generally ensured better performances because of its absolute optimality. Between the two scintillation mixtures used in CTF, based respectively on PC and PXE, the former (which is that used in the Borexino detector) exhibited the better intrinsic discrimination capabilities, both in the laboratory measurements and in the CTF set-up.

The best result in the low energy region of special interest for Borexino, obtained for events selected in the core of the CTF vessel via a 60-cm radial cut in the case of the PC scintillator and through the exploitation of Gatti's technique, gives a limited $\alpha$ tagging inefficiency of only 1.5%, while imposing a $\beta$ detection efficiency of 95%. This outcome is extremely promising and encouraging in view of the imminent start-up of the solar neutrino runs of the Borexino detector, where the PSD technique will play a key role in disentangling the true neutrino signals from the $\alpha$ background due to the natural radioactivity.

**ACKNOWLEDGMENTS**

This work has been supported in part by the Istituto Nazionale di Fisica Nucleare, the Deutsche Forschungsgemeinschaft (DFG, Sonderforschungsbereich 375), the German Bundesministerium für Bildung und Forschung (BMBF), the Maier-Leibnitz-Laboratorium (Munich), the Virtual Institute for Dark Matter and Neutrino Physics (VIDMAN, HGF), and the U.S. National Science Foundation under grants PHY-0201141, PHY-9972127, and PHY-0501118.